\documentclass[twocolumn,secnumarabic, nobibnotes, nofootinbib, aps, pre,superscriptaddress]{revtex4-2}
\usepackage{color}
\usepackage{xspace}
\usepackage{enumerate}
\usepackage{amsmath,amssymb}
\usepackage{graphicx}
\usepackage{booktabs}
\usepackage{float}
\usepackage{amsmath}
\usepackage{xcolor}
\usepackage{makecell}
\usepackage{url, changes, todonotes}
\usepackage[colorlinks=true,linkcolor=blue,citecolor=blue,urlcolor=blue,anchorcolor=blue]{hyperref}

\def\la{\langle}
\def\ra{\rangle}

\begin{document}

\title{Universal Features in Atmospheric Particulate Matter Dynamics}%

%\title{Universal Features in the Stochastic Dynamics of Atmospheric Particulate Matter}%

\author{Suchismita Banerjee}%
\email{suchib@bose.res.in}
\affiliation{S. N. Bose National Centre for Basic Sciences, Kolkata, India}

\author{Koyena Ghosh}%
\email{iamkoyenaghosh@gmail.com}
\affiliation{Indian Statistical Institute, Kolkata, India}

\author{Urna Basu}%
\email{urna@bose.res.in}
\affiliation{S. N. Bose National Centre for Basic Sciences, Kolkata, India}

\author{Banasri Basu}%
\email{sribbasu1@gmail.com}
\affiliation{Indian Statistical Institute, Kolkata, India}
%\date{December 2018}%

\begin{abstract}
\noindent We study statistical properties of  atmospheric particulate matter fluctuations using  six years  of daily PM$_{2.5}$ concentration data from fifty-four Indian cities. Despite diverse urban settings and heterogeneous climatic conditions, we find that the fluctuations show strikingly universal behaviour in both the distributional properties and temporal dynamics. 
After removing slow trends and seasonal components, the rescaled probability density functions of the residual fluctuations  collapse onto a single curve and are well described by an exponentially modified Gaussian distribution. The rescaled residual time-series for all the cities further exhibit certain robust dynamical features, with similar decay of auto-correlation functions, and power spectral densities displaying a $\sim 1/f$ decay at the tails. Finally, we propose a minimal stochastic model for the residual dynamics, which explains the observed universal features---the stationary distribution, temporal correlation, and spectral scaling. 
\end{abstract}

\maketitle
%\tableofcontents

\section{Introduction}

Atmospheric particulate matter, particularly fine particles with aerodynamic diameter less than $2.5\,\mu$m (PM$_{2.5}$), exhibits irregular and intermittent fluctuations arising from emissions, chemical transformations, meteorology and turbulent transport~\cite{seinfeld_2016}. These fluctuations reflect a complex system, where nonlinear interactions across multiple spatial and temporal scales generate variability that defies traditional deterministic or linear statistical descriptions~\cite{Kelp_2022,Beck_2020,He_2022}.
Moreover, variability in  PM$_{2.5}$ time series, manifested in amplitude and temporal structure driven by local structure and meteorology, poses challenges for unified description and motivates the identification of underlying common patterns across disparate systems.

From the perspective of nonlinear dynamics and statistical physics, identification of $universal$ $features$, shared across diverse systems irrespective of underlying microscopic details, provides a powerful framework for understanding complexity in natural systems. Universality and scaling concepts have been successfully applied to a wide range of real-world complex and  non-equilibrium systems, including turbulence~\cite{dubrulle2022multi,donzis2020universality}, finance~\cite{stanley2002scale,sarkar2018scaling}, network dynamics~\cite{macedo2017universality}, urban systems~\cite{bettencourt2020demography,Ghosh_2019,youn2016scaling}, and diverse environmental and geophysical systems~\cite{LaPorta2024,Dodds1999,smyth2019self,Entropy2021,Abbasi2024,Davidsen2007}. Equally importantly, universality is central to stochastic models  describing noise-driven complex systems, including multiplicative-noise processes, nonlinear Langevin equations, and systems exhibiting heavy-tailed distributions and $1/f$-type spectral behavior~\cite{Eliazar_2010, Yamamoto_2014, iyerbiswas_2014}.
This broad success of universality in characterizing complex non-equilibrium systems raises a natural question: given the substantial heterogeneity in emissions and meteorology, do PM$_{2.5}$ fluctuations share common stochastic features that transcend location-specific details?

Although environmental research has extensively documented  the health and atmospheric impacts of PM$_{2.5}$~\cite{world2021global,Pope01062006,burnett2018global,KG_2026}, relatively few studies~\cite{Mishra_2021, Beck_2020, Bran_2024, ShiLiuHuang2015} have examined  detailed fluctuation statistics, scaling properties, and potential universality in PM$_{2.5}$ dynamics across geographically diverse urban settings. 
To address this limitation, we investigate whether universality and signatures of scaling emerge in PM$_{2.5}$ fluctuations across multiple cities in India.

Using six years of daily PM$_{2.5}$ concentration data from fifty-four Indian cities, we demonstrate that the empirical probability density functions (PDFs) of the full time series collapse onto a common curve under appropriate rescaling, despite pronounced heterogeneity in emissions and meteorological conditions.

To probe the intrinsic stochastic dynamics, we next isolate short-term  fluctuations by removing slow seasonal and baseline components from the original time series.  The resulting residuals exhibit a striking data collapse: their rescaled  PDFs from most cities converge  onto a {\it{single universal curve}}, well described by an exponentially modified Gaussian indicative of asymmetric fluctuations and intermittency.

We further characterize the temporal structure of the residual fluctuations through auto-correlation functions (ACFs) and power spectral densities (PSDs).
The ACFs reveal persistent correlations with a power-law decay
(long-range correlations with power-law decay) over short to intermediate lags, while the PSDs display a robust scaling regime $S(f)\sim 1/f $. This behavior is characteristic of scale-invariant dynamics and is reminiscent of $1/f$ noise observed in a wide class of complex stochastic systems~\cite{Eliazar_2010, Lowen_1993}.

Finally, we propose a minimal stochastic differential equation (SDE) model that reproduces the observed residual distributional shape, correlation decay, and spectral scaling. The model provides a compact dynamical description of the PM$_{2.5}$ fluctuations and links the empirical findings to well-studied universality classes of noise-driven nonlinear systems.

These results demonstrate that urban PM$_{2.5}$ fluctuations exhibit robust and consistent statistical and dynamical signatures across diverse cities, suggesting  the existence  of universal features in their underlying stochastic dynamics. More broadly, our findings illustrate how tools from statistical physics and stochastic dynamical systems can uncover the unifying patterns in complex environmental time series.

The remainder of the paper is organized as follows: Section~\ref{sec2} describes the data compilation, processing procedures and the preliminary statistical analysis of the daily PM\textsubscript{2.5} series. In Section~\ref{sec4}, we investigate the probability distributions of daily PM\textsubscript{2.5} across cities and demonstrate universality  through collapse of the rescaled probability density functions (PDFs).
Section~\ref{sec:uni_residual} analyses residual fluctuations and their universal features through PDFs, auto-correlation functions, and power spectral densities.
In Section~\ref{sec7} we propose a (an effective) stochastic dynamical modeling framework for residual fluctuations, linking the observed statistics to the underlying dynamics. 
Finally, Section~\ref{sec8} concludes with a summary of the main findings and their implications for air-quality modeling. 

%%%%%%%%%%%%%%%%%%%%%%%%%%%%%%%%%%%%%%%%%%%%%%%%%%%%%%
\begin{figure}[t]
    \centering   \includegraphics[width=\columnwidth]{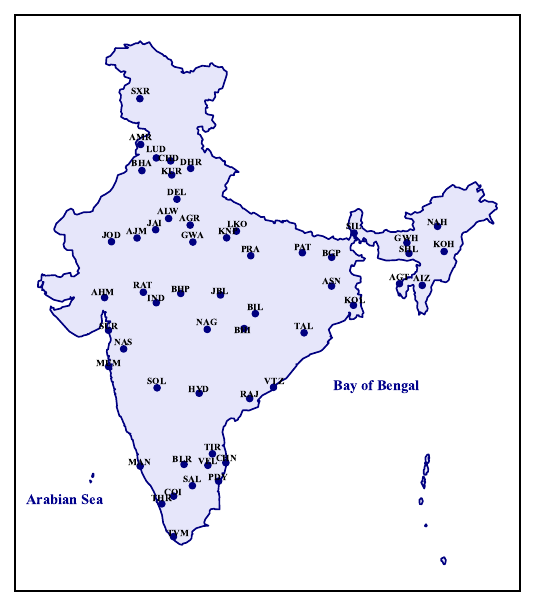}
    \caption{Map of India indicating the locations of the cities used in this analysis}
    \label{fig:map}
\end{figure}
%%%%%%%%%%%%%%%%%%%%%%%%%%%%%%%%%%%%%%%%%%%%%%%%%%%%%%
 
\section{Data Compilation and Preliminary Analysis}\label{sec2}

\noindent {\it{Data Compilation}}:
For this study, the data is collected from the Central Pollution Control Board (CPCB) of India \cite{cpcb}. The data set comprises daily average concentrations of PM\textsubscript{2.5}, $\rho(t)$, recorded at multiple monitoring stations ($N_\alpha$) located in various cities throughout India. In total, the data is  collected from 187 monitoring stations spread across 54 cities [see Fig.~\ref{fig:map} and Table~\ref{city_list}]. The time span of the dataset ranges from December 2018 to February 2025.  
Data has been collected based on measurement availability at individual monitoring station, which varies across locations and time periods.
Nevertheless, the selected cities provide broad geographic coverage across India and represent areas with a wide range of population sizes. 

%%%%%%%%%%%%%%%%%%%%%%%%%%%%%%%%%%%%%%%%%%%%%%
\begin{figure}[t]    \includegraphics[width=0.44\textwidth]{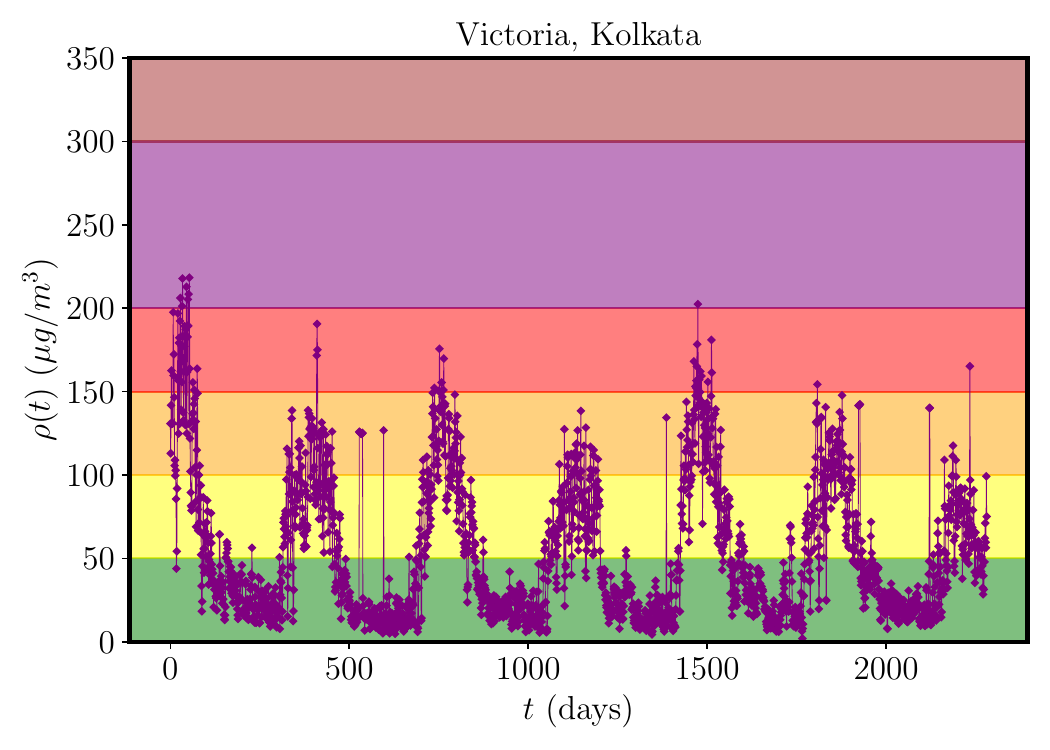}
\includegraphics[width=0.44\textwidth]{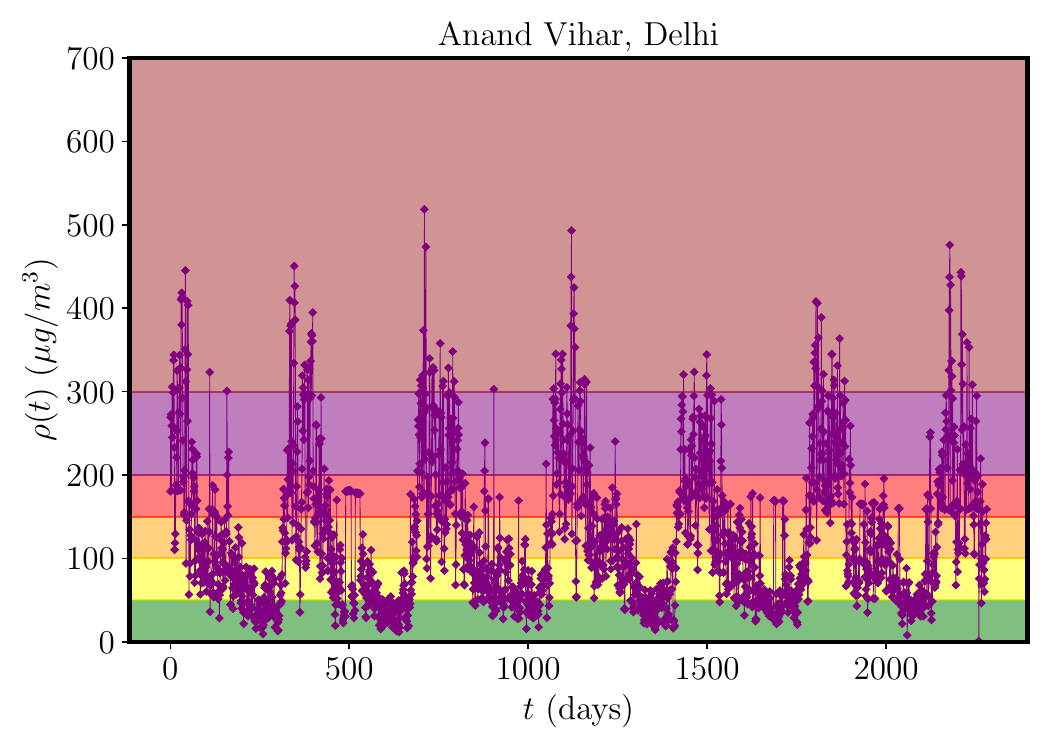}
\caption{Temporal Variation of daily PM\textsubscript{2.5} concentrations of two representative monitoring stations over the study period. Each color in the background corresponds to a specific level of health concern, with the highest level designated as hazardous~\cite{airnow_aqi_basics}.}\label{fig:ts}
\end{figure}
%%%%%%%%%%%%%%%%%%%%%%%%%%%%%%%%%%%%%%%%%%%%%%

\noindent {\it Data Preprocessing}:
The dataset contained missing values (on  average, 6.54\% of daily PM\textsubscript{2.5} observations across 54 cities.), outliers, and noisy measurements. To address these issues we employed Adaptive Kalman Filtering (AKF), an enhanced variant of the standard Kalman Filter~\cite{gibbs2011advanced}, for imputation and smoothing of time-series data\footnote{Sensitivity analysis with uniformly small root mean square error (RMSE)~\cite{rmse1}, indicates  robustness of AKF for handling missing values of PM\textsubscript{2.5}.}. 
AKF dynamically updates both the process  and measurement noise covariance matrices to accommodate changing data characteristics over time~\cite{akf} and incorporates Mahalanobis-distance-based outlier detection. This approach smooths the data while preserving its inherent temporal structure and variability of the atmospheric PM\textsubscript{2.5} time series. \\ 

\noindent{\it Temporal Trend and Seasonality}:
We now analyze the time-series data for PM\textsubscript{2.5} concentrations, $\rho(t)$.  Fig.~\ref{fig:ts} presents the daily $\rho(t)$,  plotted against time $t$ over a six-year period for two representative monitoring stations from two different cities. 
Both plots display distinct periodic behavior, illustrating a recurring pattern in each year where $\text{PM}_{2.5}$ concentrations gradually rise, reach a peak during specific months, and then decline to lower levels. This cyclic pattern repeats consistently throughout the  six-year period, reflecting a stable and predictable seasonal cycle. In addition,  several sharp and abrupt spikes are evident, signifying short-term pollution episodes or extreme events.

We also compute the mean $\langle \rho \rangle$ and standard deviation $s$ of the PM\textsubscript{2.5} concentrations\footnote{The mean and standard deviation are calculated by concatenating data from all stations within each city.} to summarize the central tendency and variability of the data (detailed values are provided in Table~\ref{tab:mean_sd_entropy_all}). $\langle \rho \rangle$ reflects the long-term pollution level, while $s$ quantifies the amplitude of concentration fluctuations.

\section{Probability Distributions of Daily PM$_{2.5}$ Concentration}\label{sec4} 

To characterize the PM\textsubscript{2.5} concentration fluctuations we first construct the probability density function $P(\rho)$ of the daily PM\textsubscript{2.5} concentrations of each city separately.  To extract $P(\rho)$ for a city, we use the entire data set of all  monitoring stations in that city and first construct the corresponding histogram. The probability density function is then obtained by normalizing the histogram, so that  $\int_0^\infty d\rho P(\rho) = 1$. For each city, the bin width is set to 10 to obtain a smooth curve. Note that, $P(\rho)$ has a dimension of $(\mu g/m^{3})^{-1}$.

\begin{figure}[t]
    \centering  \includegraphics[width=0.42\textwidth]{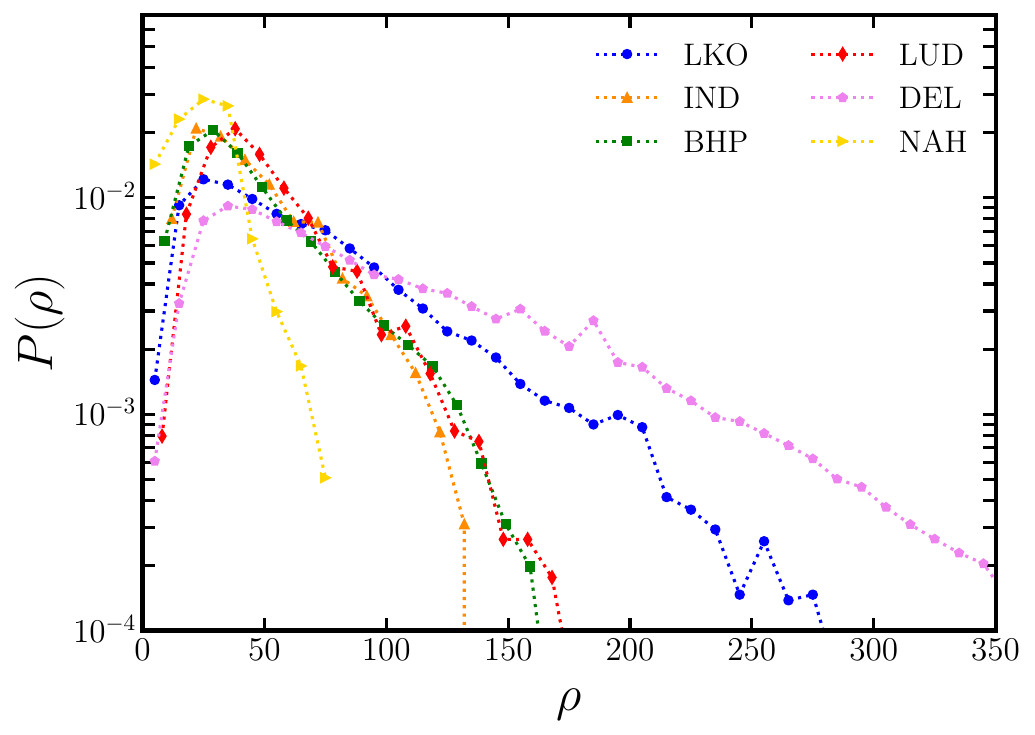}  \includegraphics[width=0.42\textwidth]{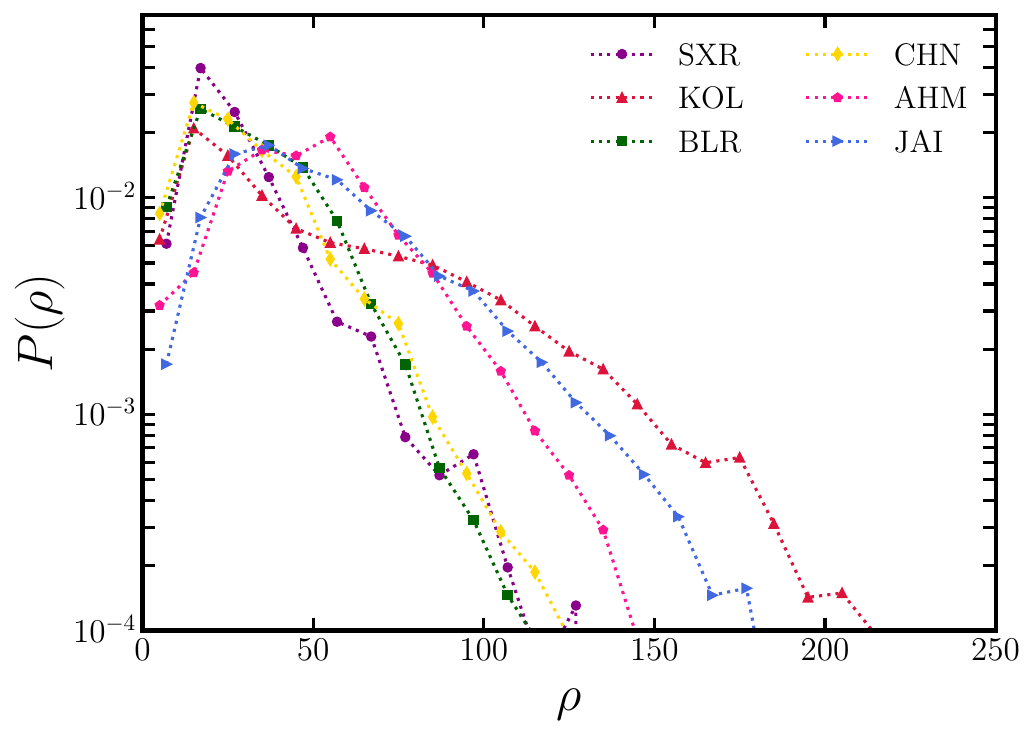}     
    \caption{Probability density functions of daily PM\textsubscript{2.5} concentration $\rho$ for various cities.}
    \label{fig:dist}
\end{figure}

Figure~\ref{fig:dist} shows plots of $P(\rho)$ versus $\rho$ for twelve representative cities. It appears from the figure $P(\rho)$ is single peaked, with strong positive skewness and has an almost exponential decay at the right tail, although the typical decay scale is different for each city. These features are shared by all the fifty four cities (not shown), and the decay scale varies over a large range [for example, for Delhi $P(\rho)$ decays from $0.01$ to $10^{-3}$ over a range of concentration $50$ to $350$ while for Bengaluru $P(\rho)$ decays from $0.02$ to $10^{-3}$ over a range of concentration $40$ to $100$].

%%%%%%%%%%%%%%%%%%%%%%%%%%%%%%%%%%%%%%%%%%%%%%
%%%%%%%%%%%%%%%%%%%%%%%%%%%%%%%%%%%%%%%%%%%%%

The qualitatively similar shape of the PDF of PM\textsubscript{2.5} distributions suggests that although absolute pollution levels vary, the distributions of PM\textsubscript{2.5} concentrations in different cities may share a similar underlying structure. To investigate this, we first shift and rescale the data for each city according to, 
\begin{align}
    \tilde \rho =\frac 1s  (\rho - \la \rho \ra)
\end{align}
where, $\la \rho \ra$ and $s$ denote the average and standard deviation of the daily  PM\textsubscript{2.5} concentration in the corresponding city. Thus, $\la \tilde \rho \ra =0$ and $\la \tilde \rho^2 \ra=1$ for all cities. Now, we construct the PDF $P(\tilde \rho)$ of the rescaled data and compare the same for all the cities\footnote{For notational simplicity we use the same letter $P$ to denote the PDF of all the variables}.

\begin{figure}[t]
\centering
\includegraphics[width=\columnwidth]{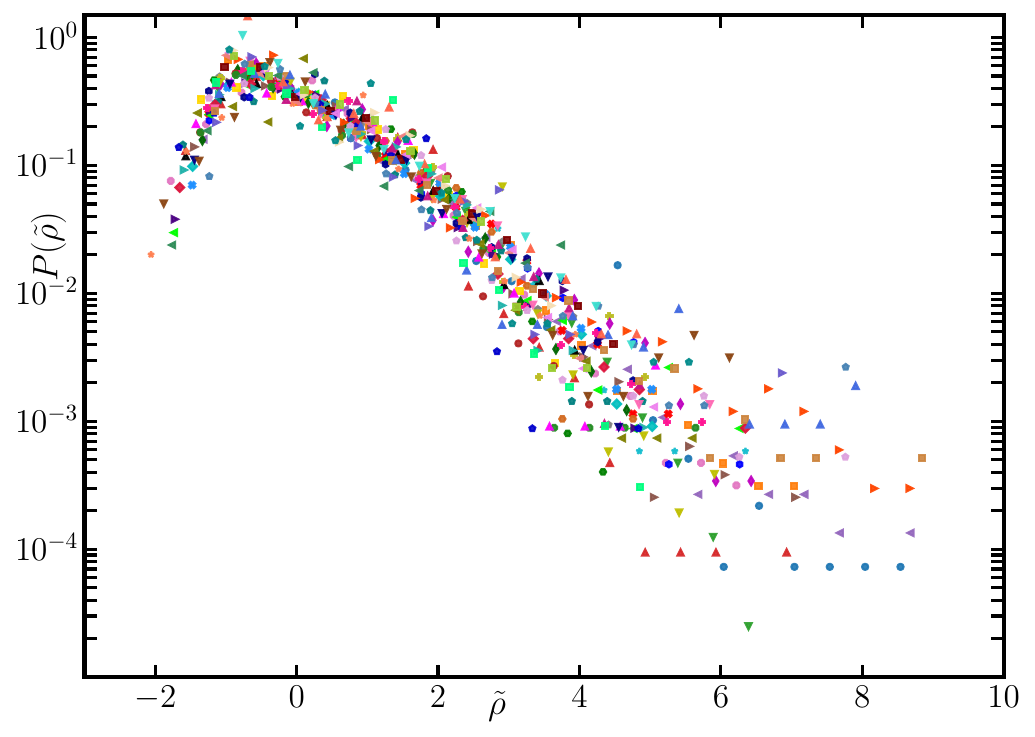}
    \caption{Data collapse of the probability density function of the rescaled PM\textsubscript{2.5} $\tilde \rho$ of all the fifty-four cities.}    \label{fig:collapse_rescaled}
\end{figure}

We find that $P(\tilde \rho)$ of the rescaled PM\textsubscript{2.5} concentrations for all the  fifty four cities almost collapse onto one another, as shown in Fig.~\ref{fig:collapse_rescaled}.
This data collapse suggests that, despite diverse local conditions, the PM\textsubscript{2.5}  fluctuations share some underlying universal features across a wide range of urban settings.

An obvious and natural question that arises here is how to understand and characterize the  universal features of PM\textsubscript{2.5} fluctuations. In particular, whether it is possible to identify the essential elements of the underlying dynamics. However, it is hard to answer this question with the bare time-series since it is non-stationary due to the presence of strong seasonal variations and long-term trends. To this end, in the following, we remove the seasonality and long-terms trends and focus on the residual data to investigate the underlying universal features.
%%%%%%%%%%%%%%%%%%%%%%%%%%%%%%%%%%%%%%%%%%%%%%%%%
%%%%%%%%%%%%%%%%%%%%%%%%%%%%%%%%%%%%%%%%%%%%%%%%%

\section{Universality in Residual Fluctuations}\label{sec:uni_residual}
A non-stationary time series is typically modeled as a combination of three components: a long-term trend, a seasonal or cyclical component, and a short-term residual component. In this section, we decompose the observed time series and characterize the residual PM$_{2.5}$ fluctuations in terms of  their statistical, temporal, and spectral properties.

We begin by decomposing the observed time series using a multiplicative formulation~\cite{multiplyts}, which is well suited for pollutant data, as the amplitude of seasonal variations typically scales with the overall concentration level. Under this framework,  the residual component $R_t$  is obtained by removing the long-term trend $T_t$, cyclical component $S_t$ from the original time-series $\rho_t$ for each station. 

%%%%%%%%%%%%%%%%%%%%%%%%%%%%%%%%%%%%%%%%%%%%%%%%%%%
\begin{figure}[tbh]
\centering
\includegraphics[width=\columnwidth]{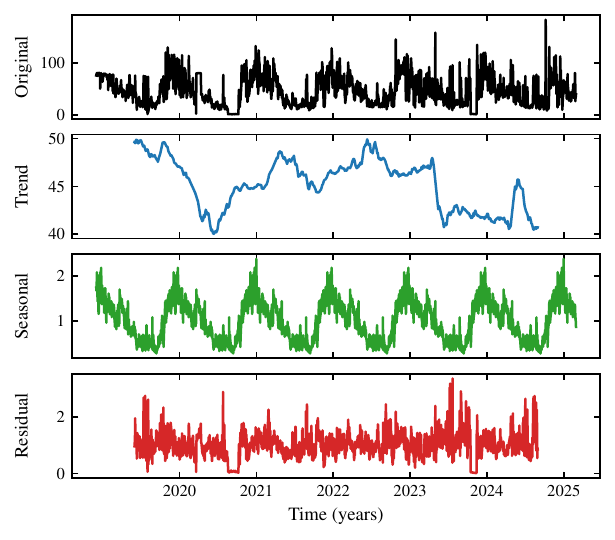}
    \caption{Decomposition of original PM\textsubscript{2.5} series into trend, seasonality and residuals using method of moving average. Here we show the data from the monitoring station Sanathnagar in HYD.}
  \label{fig:decomposition}
\end{figure}
%%%%%%%%%%%%%%%%%%%%%%%%%%%%%%%%%%%%%%%%%%%%%%%%%%%%%%

Given the pronounced annual cycle exhibited in the original PM\textsubscript{2.5} time series [see Fig.~\ref{fig:ts}], to isolate the short-term fluctuations, we estimate the combined trend--seasonal component using a $k$-period moving-average method with $k=365$. The residuals are then defined as
\begin{equation}
R_t = \frac{\rho_t}{\widehat{T}_t\,\widehat{S}_t},
\end{equation}
where $\widehat{T}_t$ and $\widehat{S}_t$ denote moving-average estimates of the trend and seasonal components, respectively. This procedure effectively removes long-term and seasonal variability while preserving short-time dynamics. Details of the time series decomposition is provided in  Appendix~\ref{ap:decomp}. 

Figure~\ref{fig:decomposition} illustrates the decomposition for a representative monitoring station; qualitatively similar behavior is observed across all stations. The decomposed series reveals a time-varying, non-linear trend with visible structural changes over time. In contrast, the seasonal component shows a strong and stable pattern, characterized by regular and consistent periodic fluctuations. 
The residual component contains the short-term stochastic fluctuations and is statistically stationary, thereby enabling a detailed investigation of potential universal patterns in the intrinsic stochastic variability and  statistical structures of PM$_{2.5}$. In the following, we focus exclusively on the properties of the residual series $R_t$ and examine their statistical characteristics across different cities to explore potential universality.

\subsection{Stationary distribution of residual time-series}\label{sec:stationary}

We start by taking a closer look at the residual time series. From Fig.~\ref{fig:decomposition} it appears that the residual series is stationary and consists of small fluctuations about an average value with occasional large kicks.  To characterize the nature of this residual component $R_t$ for each city, as before,  we aggregate the data from all the stations.  To facilitate comparison we shift and scale the data for each city following,
\begin{align} 
\tilde R = \frac {1}{s_R} (R - \langle R \rangle), 
\end{align}
where $\la R \ra$ and $s_R$ denote the average of the residual dataset and  the corresponding standard deviation for that city, respectively.  This rescaling ensures that for each city, we now have, $\la \tilde R \ra =0 $ and $\la \tilde R^2 \ra =1$. 

%%%%%%%%%%%%%%%%%%%%%%%%%%%%%%%%%%
\begin{figure}[t]
    \centering   \includegraphics[width=\columnwidth]{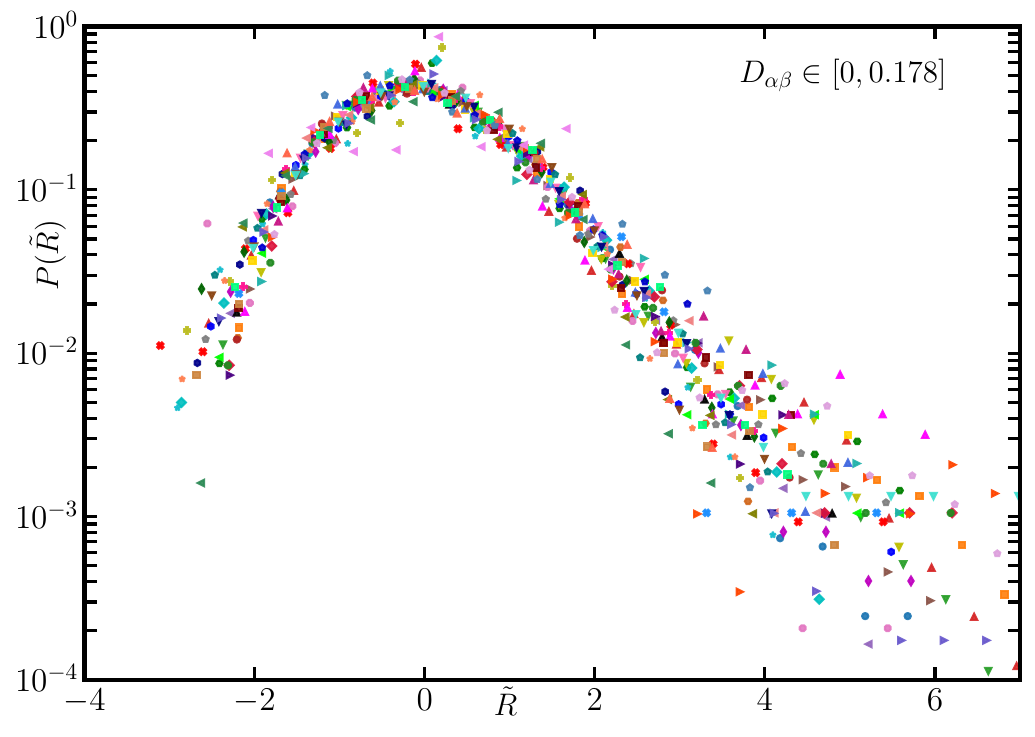}
    \caption{Collapse of the empirical PDF $P(\tilde R)$ of rescaled PM\textsubscript{2.5} residuals $\tilde R$ of fifty one cities belonging to a single cluster with pairwise $D_{\alpha \beta}$s  ranging from $0$ to $0.178$.}    \label{fig:collapse_residual}
\end{figure}
%%%%%%%%%%%%%%%%%%%%%%%%%%%%%%%%%%%%%%%
Using this empirical rescaled data, we construct the PDFs $P_\alpha(\tilde R)$ for the $\alpha$-th city with $\alpha=1, \cdots 54$. Our goal is to investigate whether any universal pattern governs the pollutant fluctuations. To this end, we quantify the similarity among the residual distribution of different cities using Jensen-Shanon divergence.

The Jensen-Shannon divergence (JSD)~\cite{js} between two probability distributions $P_{\alpha}(\tilde R)$ and $P_{\beta}(\tilde R)$ is defined as,
\begin{equation}
    D_{\alpha \beta}(P_\alpha||P_\beta) =\int \frac{d \tilde R}2\,\Bigg [P_\alpha \ln \frac{P_\alpha}{M_{\alpha \beta}} + P_\beta(\rho) \ln \frac{P_\beta}{M_{\alpha \beta}} \Bigg],
 \end{equation}
where $M_{\alpha \beta}(\tilde R) = \frac{1}{2}[P_{\alpha}(\tilde R)+P_{\beta}(\tilde R)]$ denotes the average of $P_{\alpha}(\tilde R)$ and $P_{\beta}(\tilde R)$. JSD is nothing but a symmetrized version of Kullback-Leibler divergence~\cite{KL} and  provides a quantitative measure of the degree of similarity between two distributions. Clearly, the JSD for two identical  distributions vanishes. In fact, it is easy to show that JSD is always non-negative and is bounded in the regime $D_{\alpha \beta} \in  [0, \ln 2 \simeq 0.693]$ with a lower value indicating a higher degree of similarity.

We compute the pairwise Jensen--Shannon divergence (JSD) values $D_{\alpha \beta}$ for all city pairs  $ (\alpha, \beta)$ [a total of $^{54}C_2=1431 $ pairs] and find that all $D_{\alpha \beta}$ are generally very low, ranging from $0$ to $0.205$. This indicates that the residual distributions of all the cities are similar. To characterize this similarity further, we group cities based on their $D_{\alpha \beta}$ values using hierarchical clustering \cite{nielsen2016hierarchical,hierarchical_clustering1}. 
This unsupervised clustering approach initializes each observation (here, the cities) as a distinct  cluster and iteratively merges the most similar pairs based on their corresponding $D_{\alpha \beta}$ values, yielding a group structure inherent in the data. Interestingly, we find that the hierarchical clustering on the $D_{\alpha \beta}$ values results in a two-cluster partition of the cities. The larger cluster comprises fifty-one cities ($\sim$ 94.4\% of the data), with pairwise $D_{\alpha \beta}$s  ranging from $0$ to $0.178$.  Three  smaller cities (Salem, Tirupati, Surat) likely reflects  deviations from the universal pattern due to smaller sample sizes and limited data availability. 

Figure~\ref{fig:collapse_residual} shows a plot of the distribution of the scaled residual fluctuations $\tilde R$ for the fifty-one cities which show an extremely good data collapse. This observed collapse suggests that despite substantial differences in geography, population density, and emission sources, the short-term fluctuations of the PM$_\text{2.5}$ show a universal underlying statistical behaviour. Identifying this universality is one of the main results of this work.

%%%%%%%%%%%%%%%%%%%%%%%%%%%%%%%%%%%%%%%%%%%%%%%%%%
\begin{figure}[t]
    \centering  
    \includegraphics[width=0.48\textwidth]{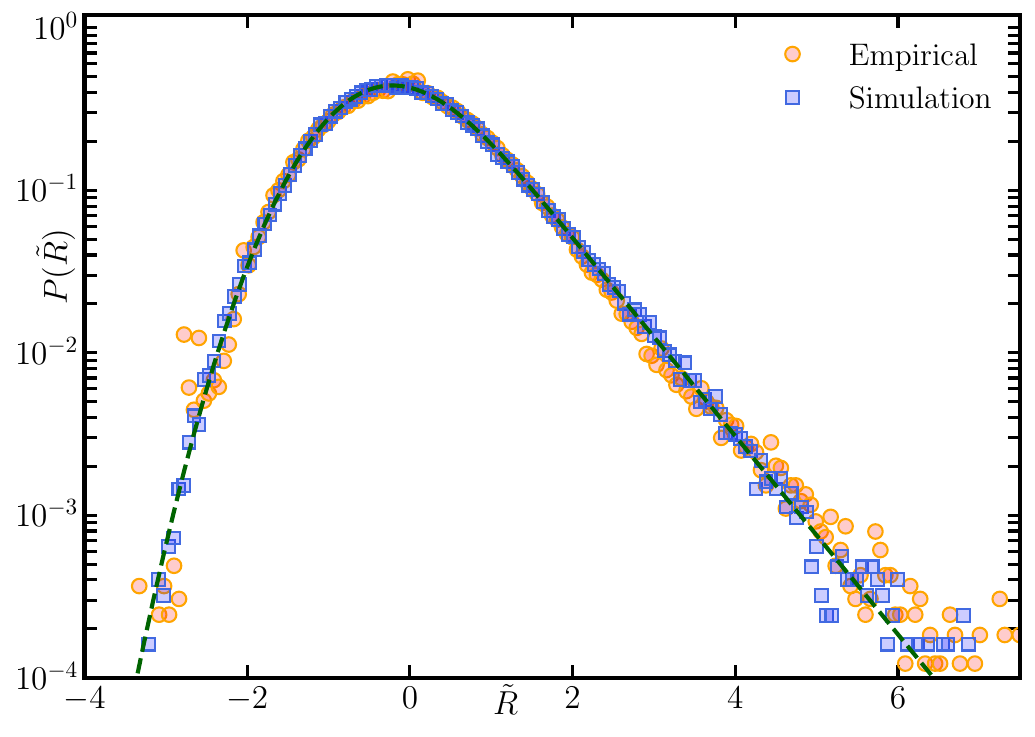}
    \caption{Plot of probability distribution function $P(\tilde{R})$ vs $\tilde{R}$ of residual PM\textsubscript{2.5}. The orange circles indicate the universal scaling curve obtained from the empirical data and blue squares indicate the data obtained from numerical simulations using Eq.~\eqref{eq:X_decomp}. The dashed line shows the EMG curve with with $\mu=-0.706$, and hence $\sigma=0.708$ and $\lambda=1.416$.}
  \label{fig:pdfsimulation}
\end{figure}
%%%%%%%%%%%%%%%%%%%%%%%%%%%%%%%%%%%%%%%%%%%%%%%%%%%%%

%%%%%%%%%%%%%%%%%%%%%%%%%%%%%%%%%%%%%%%%%%%%%%%%%%%%
% \begin{figure}[t]
%     \centering   \includegraphics[width=\columnwidth]{PRE/PDF.pdf}
%     \caption{\added{The universal scaling curve obtained from the aggregated data collapse of residual PDFs (symbols). The dashed line shows the} EMG function given by Eq.~\eqref{eq:emg_f} with $\mu=-0.706$, and hence $\sigma=0.708$ and $\lambda=1.416$.}   \label{fig:fit_universal}
% \end{figure}
%%%%%%%%%%%%%%%%%%%%%%%%%%%%%%%%%%%%%%%%%%%%%%%%%%%%%%

Motivated by this observation, we aggregate the data from all the fifty-one cities to construct a single representative curve that displays 
the shared statistical behavior of PM$_\text{2.5}$ residuals. This is shown in  Fig.~\ref{fig:pdfsimulation} with orange circles.  We refer to this combined empirical distribution as  the ``universal scaling curve". To obtain a compact and tractable representation of the universal scaling curve, it is useful to have an analytical form for this empirical scaling function.  From Fig.~\ref{fig:collapse_residual} it is clear that this universal distribution exhibits clear deviations from Gaussian behavior: it is positively skewed, with a broad Gaussian-like peak and an exponential-like tail indicating the presence of large fluctuations and intermittency. This shape motivates us to propose an ansatz that the universal scaling function for $\tilde R$ is given by an exponential modified Gaussian distribution,
\begin{align}
F(\tilde{R})=\frac{\lambda}{2} \, \exp{\left[ \frac{\lambda \sigma^2}{2} + \lambda(\mu - \tilde R) \right]}\, \text{erfc}\left[
\frac{\lambda\sigma^{2}+\mu-\tilde{R}}{\sqrt{2}\,\sigma}
\right]. \label{eq:EMG}
\end{align}
Here $\text{erfc}(u)$ denotes the complimentary error function and $\mu, \lambda$ and $\sigma$ are three real parameters which characterize this distribution function.
To determine these parameters for our case, we first note that for the rescaled residual data, we have $\la \tilde R \ra=0$ and  $\la \tilde R^2 \ra =1$ which implies that [see Appendix~\ref{app:EMG} for more details],
\begin{equation}
\lambda = -\frac{1}{\mu}, \qquad \sigma^2 = 1 - \mu^2.
\end{equation}
Finally, $\mu$ can be determined from estimating the skewness of the empirical distribution,
\begin{equation}
\mu = -\left(\frac{\langle \tilde{R}^3 \rangle}{2}\right)^{1/3}. \label{eq:mu}
\end{equation}
Equation~\eqref{eq:EMG} thus reduces to a one parameter function,
\begin{align}
 F(\tilde{R})=- \frac{1}{2\mu}\exp\left[\frac 1 \mu \left(\tilde R + \frac 1{2\mu}\right)- \frac 32\right] \text{erfc} \left[\frac{2 \mu^2 - \mu \tilde R-1}{\mu \sqrt{2(1-\mu^2)}} \right]. \label{eq:emg_f} 
\end{align}
where $\mu$ is given by Eq.~\eqref{eq:mu}. 

We estimate the skewness $\la \tilde R^3 \ra$ of the universal scaling curve from the empirical data, which leads to $\mu \simeq -0.706$, and, in turn, $\sigma=0.708$ and $\lambda=1.416$. A plot of $F(\tilde{R})$ with these parameters is compared to the empirical universal scaling function in Fig.~\ref{fig:pdfsimulation} which shows an excellent match. Minor deviations are observed only in the tails, likely reflecting rare extreme pollution events. 
To quantify the goodness of fit, we also estimate the Kolmogorov-Smirnov statistic value, which turns out to be 
$\approx 0.03$, indicating that Eq.~\eqref{eq:emg_f} describes the stationary distribution of the residual fluctuations with a very good accuracy. It should be emphasized that there are no free fitting parameters here, all the three parameters are estimated from the data. 

The Exponentially Modified Gaussian distribution is widely used~\cite{ali2022comparison,Zhen_2014,Golubev_2010} to model stochastic processes when Gaussian background fluctuations are intermittently perturbed by a single exponential scale. 

To investigate the origin of the universal EMG distribution for PM$_{2.5}$ residuals, we next focus on the dynamics of residual fluctuations.

\subsection{Dynamical features of residual time-series}\label{sec:dynamic}

The emergent universal form of the stationary residual distribution indicates that the mechanisms governing the short-term time evolution of PM$_{2.5}$ may possess common, location-independent features. This points to an underlying dynamical process with shared structure across locations, beyond the observed similarity in stationary fluctuations. To probe this, we examine the dynamical features of the residual time series by extracting the auto-correlation and power spectral density for each city.

To extract the auto-correlation of residual time-series for a given city $\alpha$, we first compute, for the $i$-th monitoring station in that city, the auto-correlation at lag $\tau$,
\begin{align} 
C_{\alpha,i} (\tau) = \frac 1{T_i - \tau}\sum_{t=1}^{T_i-\tau}  \tilde R_t \tilde R_{t+\tau},
\end{align} 
where $T_i$ denotes the size of the corresponding time-series, that is, the number of daily data available for that monitoring station. Then, city-wise auto-correlation function (ACF) is obtained by averaging over all the monitoring stations in that city,
\begin{align}
C_\alpha(\tau) = \frac{1}{N_\alpha} \sum_{i=1}^{N_\alpha} C_{\alpha,i}(\tau),    
\end{align}
where $N_\alpha$ denotes the total number of monitoring stations in the $\alpha$-th city. Note that the definition ensures $C_\alpha(0)=1$. The averaging over the stations reduces station‐level noise and yields a robust estimate of the temporal correlation structure characteristic of the city as a whole.

%%%%%%%%%%%%%%%%%%%%%%%%%%%%%%%%%%%%%%%%%%%%%%%%%%%%%%%%%
\begin{figure}[t]
    \centering   \includegraphics[width=0.5\textwidth]{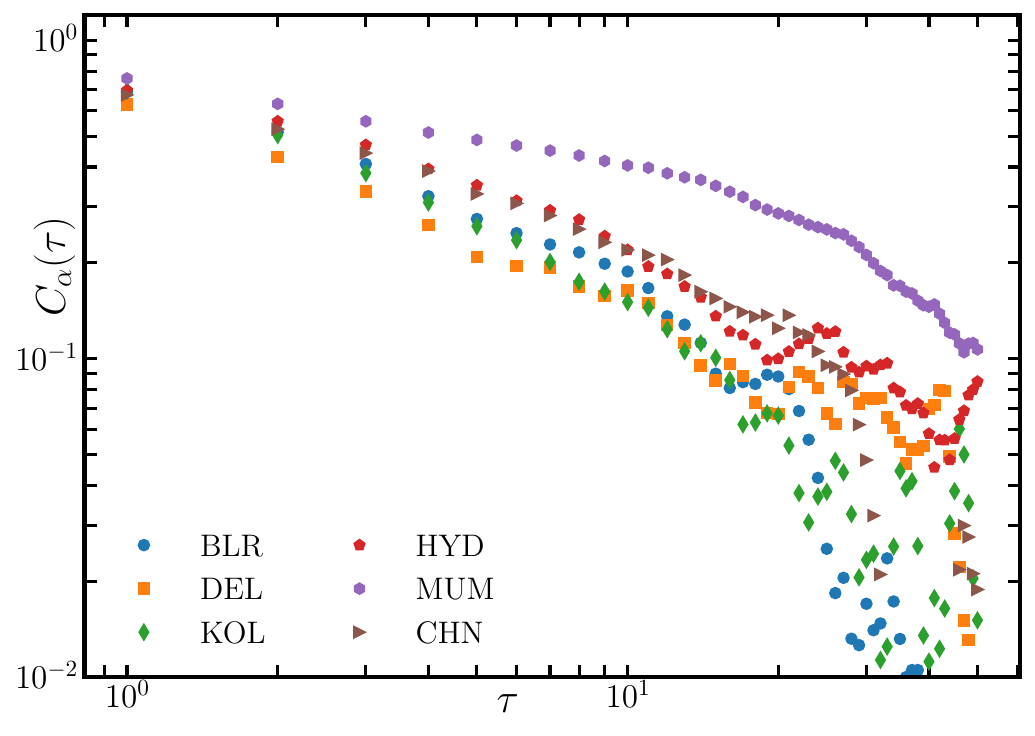}
    \caption{Plot of auto-correlation functions $C_{\alpha}(\tau)$ vs lag $\tau$ for some representative cities $\alpha$. The ACF is obtained by averaging the rescaled residual data from all monitoring stations in the corresponding cities.}
    \label{fig:ACF}
\end{figure}
%%%%%%%%%%%%%%%%%%%%%%%%%%%%%%%%%%%%%%%%%%%%%%%%%%%%%%%%%%%%

We repeat this analysis for all the fifty-four cities which reveals interestingly similar temporal pattern.  Figure~\ref{fig:ACF} shows a plot of $C_\alpha(\tau)$ vs $\tau$ for six representative cities which shows that the ACF first shows a slow decay, up to approximately $\tau \sim 10$ days, which is followed by a faster exponential-like decay. The same feature is observed for all the cities.

To further characterize the temporal dynamics in the frequency domain, we next examine the power spectral density (PSD) of the residual time series. The PSD provides a complementary perspective by revealing how fluctuations are distributed across different time scales and allows for a clearer identification of scaling behavior and dominant frequencies. To extract the PSD of the residual time-series for a given city $\alpha$, we first compute, for the $i$-th monitoring station in that city, the PSD at frequency $f$,
\begin{align}
S_{\alpha,i}(f) = \frac{1}{T_i} \left| \sum_{t=1}^{T_i} \tilde R_t \, e^{-2\pi i f t} \right|^2,
\end{align}
where $T_i$ denotes the size of the corresponding time-series. 
Since the time-series is sampled at fixed interval of $\Delta t=1$ day, the frequency $f$ is bounded in the regime $0 \le f \le 1/2$. It should be noted that $S_{\alpha,i}(f)$ is nothing but the Fourier transform of $C_{\alpha,i}(\tau)$,
\begin{align}
S_{\alpha,i}(f) = \sum_{\tau = 0}^{T_i-1} C_{\alpha,i}(\tau)\, e^{-2\pi i f \tau},
\end{align}
in accordance with the Wiener–Khinchin theorem~\cite{WK_theorem1,WK_theorem2}. In practice, we estimate the PSD using the Welch method~\cite{welch}, which reduces noise by averaging over segmented, windowed realizations of the time series.
Finally, the city-wise PSD is obtained by averaging over all the monitoring stations in that city,
\begin{align}
S_\alpha(f) = \frac{1}{N_\alpha} \sum_{i=1}^{N_\alpha} S_{\alpha,i}(f),
\end{align}
where $N_\alpha$ denotes the total number of monitoring stations in the $\alpha$-th city.

%%%%%%%%%%%%%%%%%%%%%%%%%%%%%%%%%%%%%%%%%%%%%%%%%%%%%%%%%%%%
\begin{figure}[tbh]
    \centering   \includegraphics[width=0.5\textwidth]{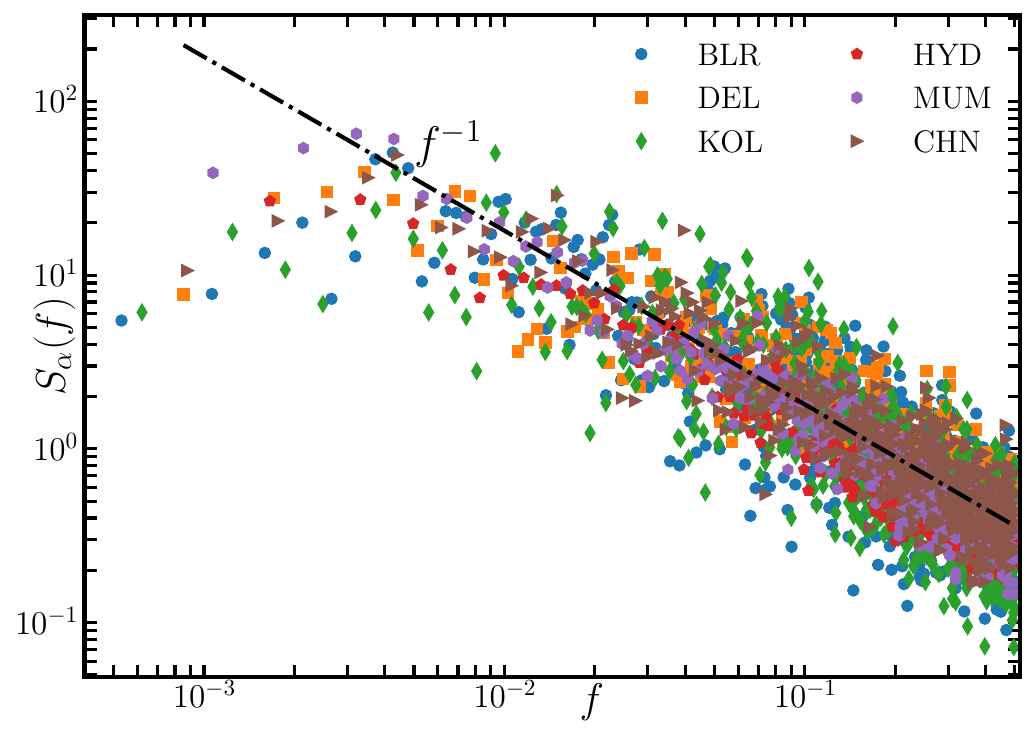}
    \caption{Plot of power spectral density $S_{\alpha}(f)$ vs frequency $f$ for some representative cities $\alpha$.  The PSD is obtained by considering the average of the rescaled residual data from all monitoring stations in the corresponding cities. The dashed line indicates the reference $1/f$ line.}
    \label{fig:PSD_mega}
\end{figure}
%%%%%%%%%%%%%%%%%%%%%%%%%%%%%%%%%%%%%%%%%%%%%%%%%%%%%%%%%%%%

Figure~\ref{fig:PSD_mega} illustrates the behaviour of the PSD for a set of representative cities including  five mega cities. It appears that the PSD for all these cities show very similar behaviour. In particular, we see, 
\begin{align}
 S_\alpha(f) \sim 1/f   
\end{align}
at the tail. The same $1/f$ decay is observed for all the cities, reinforcing the presence  of universality in the underlying dynamics.

A $1/f$ decay in the PSD is often observed in complex systems including biological and neural signals, climate records, electronic systems, and economic data and is commonly interpreted as evidence of long-range temporal correlations and multiple time-scale structures~\cite{Eliazar_2010, Keshner_1982, Weissman_1988}. Such behavior is noteworthy as it indicates scale-free dynamics with dominant low-frequency fluctuations and no single governing timescale. 
The simultaneous presence of these  universal features across cities indicates robust, scale-invariant dynamics and motivates further theoretical investigation to better understand the underlying mechanisms.

\section{Stochastic Modeling of Residual Fluctuations} \label{sec7}

The observed universality in both the stationary distribution and the dynamical fluctuations of residual PM\textsubscript{2.5} across cities naturally motivates a stochastic description in which the time evolution is driven by some effective noise around a steady average value. In this section, we propose a minimal stochastic model for the temporal evolution of PM\textsubscript{2.5} that captures the observed universal features of both the stationary distribution and the dynamical fluctuations including the $1/f$ decay of the PSD.

%%%%%%%%%%%%%%%%%%%%%%%%%%%%%%%%%%%%%%%%%%%%%%%%%%
\begin{figure*}[t]
    \centering    \includegraphics[width=0.32\textwidth]{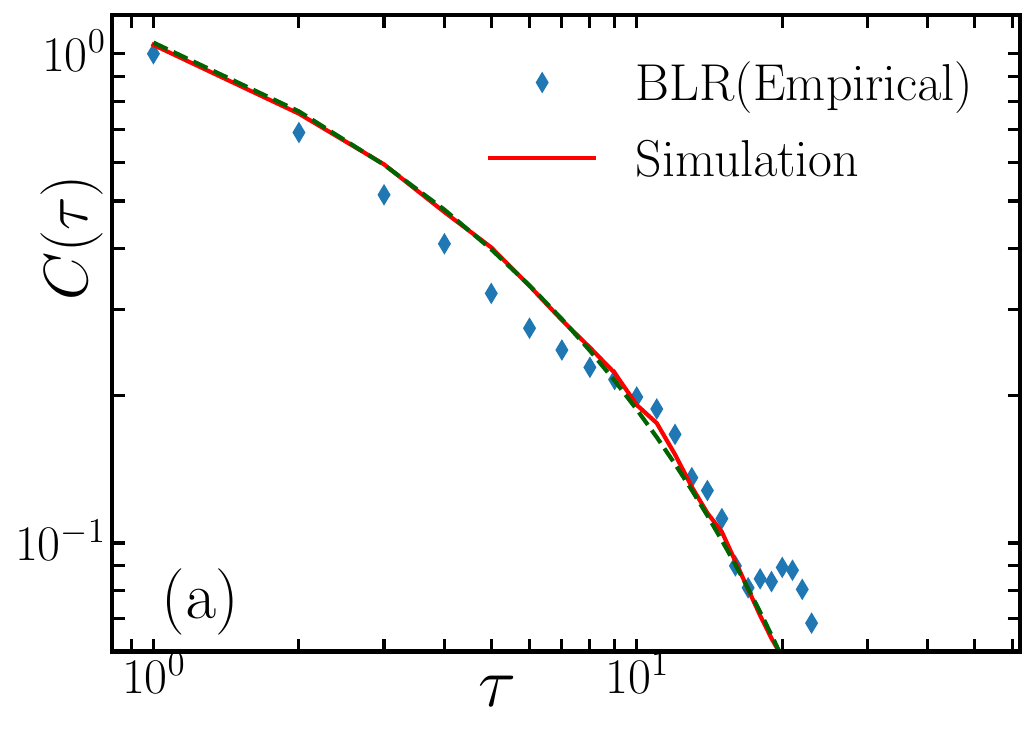}   \includegraphics[width=0.32\textwidth]{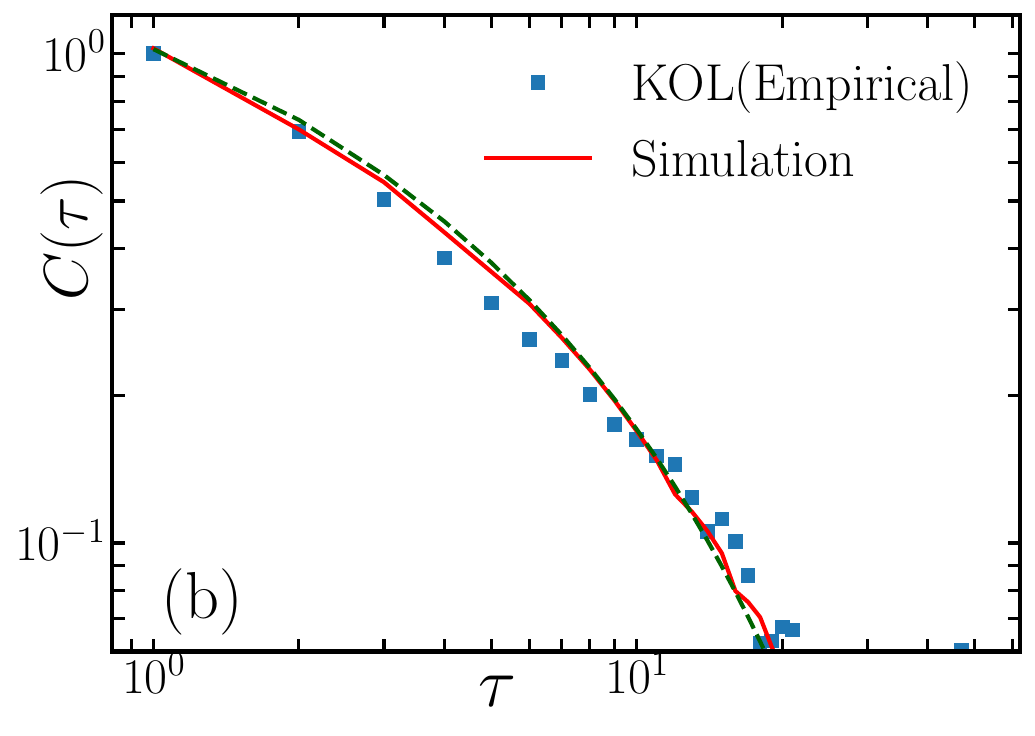}    \includegraphics[width=0.32\textwidth]{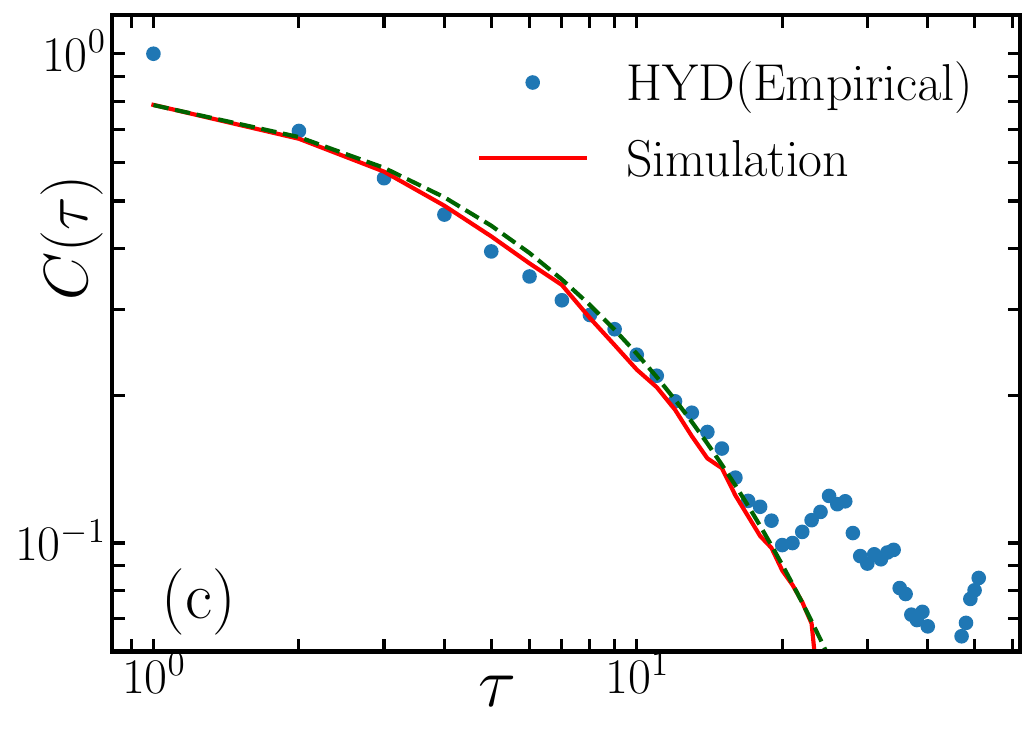}    
    \caption{Plot of auto-correlation function $C(\tau)$ vs $\tau$ for (a) BLR, (b) KOL, and (c) HYD. The symbols indicate the $C(\tau)$ extracted from the empirical data while the solid and dashed lines indicate the theoretical prediction Eq.~\eqref{eq:c_tau_theory}, and the data obtained from numerical simulations, respectively. The value of $\nu_\text{min}$ and $\nu_\text{max}$ are estimated from fitting Eq.~\eqref{eq:c_tau_theory} to the empirical data, which are given by $\nu_\text{min}=0.075$ and $\nu_\text{max}=2.0$ for BLR; $\nu_\text{min}=0.07$ and $\nu_\text{max}=2.0$ for KOL, and $\nu_\text{min}=0.056$ and $\nu_\text{max}=0.35$  for HYD. These values, in turn, are used for the numerical simulations for the respective cities.}
   \label{fig:acfsimulation}
\end{figure*}
%%%%%%%%%%%%%%%%%%%%%%%%%%%%%%%%%%%%%%%%%%%%%%%%%%

The EMG form of the stationary distribution suggests that the residual series comprises a Gaussian background with intermittent exponential fluctuations. The simplest way to model a dynamics with stationary Gaussian fluctuations is through Ornstein-Uhlenbeck (OU) process. However, a single OU process has a purely exponential auto-correlation and is thus not suitable for modeling the PM\textsubscript{2.5} fluctuations. Instead we consider a weighted sum of independent OU processes along with intermittent kicks with exponentially distributed strengths. 

We propose that the universal dynamics of the residual component of PM\textsubscript{2.5} can be described by a process,
\begin{equation}
\mathcal{R}(t) = \mu + \sigma\,G(t) + Z(t),
\label{eq:X_decomp}
\end{equation}
where $\mu$ and $\sigma$ are constants, $G(t)$ and $Z(t)$ are two independent processes, modeling, respectively, the  Gaussian background and the intermittent noise. In the following we define these two processes separately.

The stochastic process $G(t)$ refers to a weighted sum of $K$ independent OU processes,
\begin{equation}
G(t) = \sum_{i=1}^{K} w_i\,Y_i(t),~~ \text{with}~~~ \sum_{i=1}^K w_i^2=1, \label{eq:G_mixture}
\end{equation}
where $Y_i(t)$ evolves following, 
\begin{equation}
\dot Y_i(t) = -\nu_i \,Y_i(t)  + \sqrt{2 \nu_i}\,\eta_i(t),
\quad i=1,\dots,K,
\label{eq:OU}
\end{equation}
with mobility $\nu_i$. Here $\{ \eta_i(t), i=1,2, \cdots K \}$ refer to independent Gaussian white noises with $\la \eta_i(t) \eta_j(t') \ra = \delta_{ij}\delta(t-t')$. It is straightforward to see from Eq.~\eqref{eq:OU} that $Y_i(t)$ reaches a stationary state at times $t \gg \nu_i^{-1}$ with 
$\la Y_i\ra=0$ and $\la Y_i^2 \ra=1$. In fact, the stationary state is characterized by Gaussian distribution of each $Y_i$,
\begin{align}
P(Y_i) = \frac 1{\sqrt{2 \pi }} e^{-Y_i^2/2}.
\end{align} 
Thus, the sum $G(t)$ also has a Gaussian distribution in the stationary state, with mean $\la G \ra=0$ and variance $\la G^2 \ra = \sum_{i=1}^K w_i^2=1$. The parameters $K$ and $\{w_i\}$ need to be determined following this constraint; we will discuss this issue later.

Next we consider $Z(t)$ which represents independent intermittent positive fluctuations capturing irregular bursts not accounted for by the smooth background dynamics. We model this burst as an exponentially distributed random variable that acts at discrete times $\{t_n \}$. For simplicity, we take $\{t_n =1, 2 , \cdots N\}$, which also makes it easy to mimic the empirical data. Thus, $Z(t)$ is given by a discrete time series $\{z_1, z_2, \cdots z_i, \cdots\}$ where $z_i$ are independent and identically distributed random variables drawn from $p(z) = \lambda e^{-\lambda z}$.  

Finally, sampling $G(t)$ at discrete times ${t=1,2, \cdots N}$ and combining it with the discrete time noise $Z(t)$, following Eq.~\eqref{eq:X_decomp}, we get a time-series $X(t)$ which becomes stationary eventually. The fact that $X$ is a sum of a Gaussian distributed variable $G$ and an exponentially distributed variable $Z$, shifted by a constant, ensures that, in the stationary state $\mathcal{R}$ follows the EMG distribution~\cite{ali2022comparison,Zhen_2014,Golubev_2010}. Moreover, from Eq.~\eqref{eq:X_decomp}, we have,
\begin{align}
\la \mathcal{R} \ra &= \mu + \frac 1\lambda, ~\text{and} \cr 
\la \mathcal{R}^2\ra -\la \mathcal{R} \ra^2 & = \sigma^2 +  \frac 1{\lambda^2}. 
\end{align}
Thus, the stationary distribution of $\mathcal{R}$ is given by \eqref{eq:EMG} with the parameters $\mu, \sigma$ and $\lambda$. As discussed before, to model the residual series, we must have $\la \mathcal{R} \ra =0$ and $\la \mathcal{R}^2 \ra =1$, thus these parameters are related and can be extracted from the empirical data easily [see Sec~\ref{sec:stationary} above]. It should be noted that the stationary distribution is independent of $\{ \nu_i\}.$

However, our goal is not only to model the stationary distribution, but also the dynamical correlations. To this end, we first note that, from Eq.~\eqref{eq:OU}, we can write the stationary auto-correlation of $Y_i(t)$ as,
\begin{align}
\la Y_i(t) Y_j(t+ \tau) \ra = \delta_{ij} e^{-\nu_i \tau}.
\end{align}
On the other hand, since $z_i$ is independent at each time, $\la Z(t) Z(t+\tau) \ra = \la z\ra^2 = 1/\lambda^2$. Thus, the auto-correlation of the process $\mathcal{R}(t)$ is given by,
\begin{align}
 C(\tau) & \equiv \la  \mathcal{R}(t) \mathcal{R}(t+\tau) \ra \cr  & = \sigma^2 \la G(t) G(t+\tau) \ra 
 = \sigma^2 \sum_{i=1}^K w_i^2 e^{- \nu_i \tau}. \label{eq:C_tau_th}
\end{align}
The weighted sum over the different modes $Y_i$ implies that the auto-correlation is not purely exponential. The corresponding power-spectral density can be easily obtained, and is given by a weighted sum of Lorentzians with different widths,
\begin{align}\label{eq:psd}
    S(f) = 2\int_0^\infty d\tau \, C(\tau) \cos (2\pi f \tau) =2 \sigma^2 \sum_{i=1}^K \frac{w_i^2 \, \nu_i}{\nu_i^2 + 4 \pi^2 f^2}.
\end{align}

%%%%%%%%%%%%%%%%%%%%%%%%%%%%%%%%%%%%%%%%%%%%%%%%%%%%
\begin{figure*}[t]
    \centering  \includegraphics[width=0.32\textwidth]{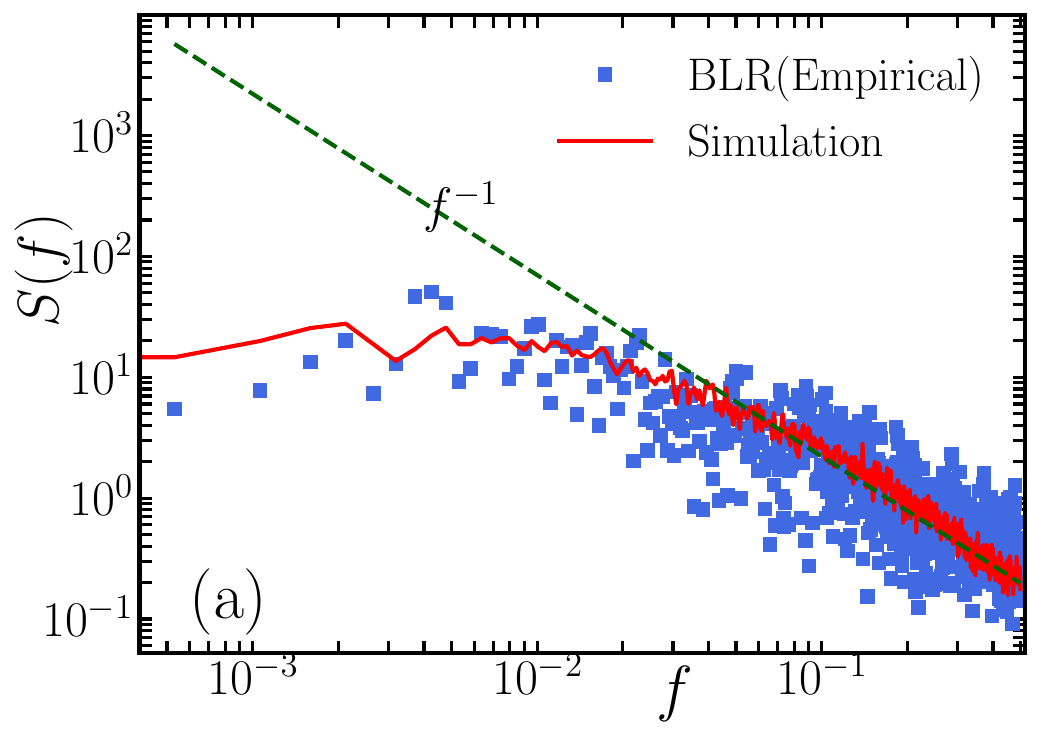}
    \includegraphics[width=0.32\textwidth]{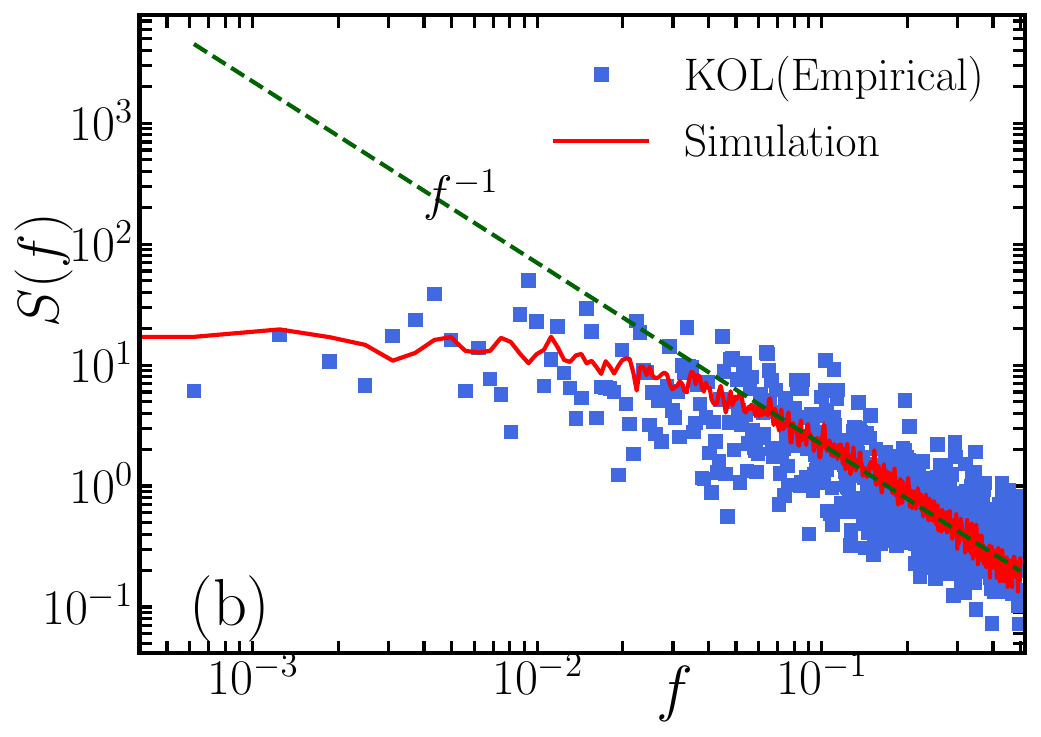}
    \includegraphics[width=0.32\textwidth]{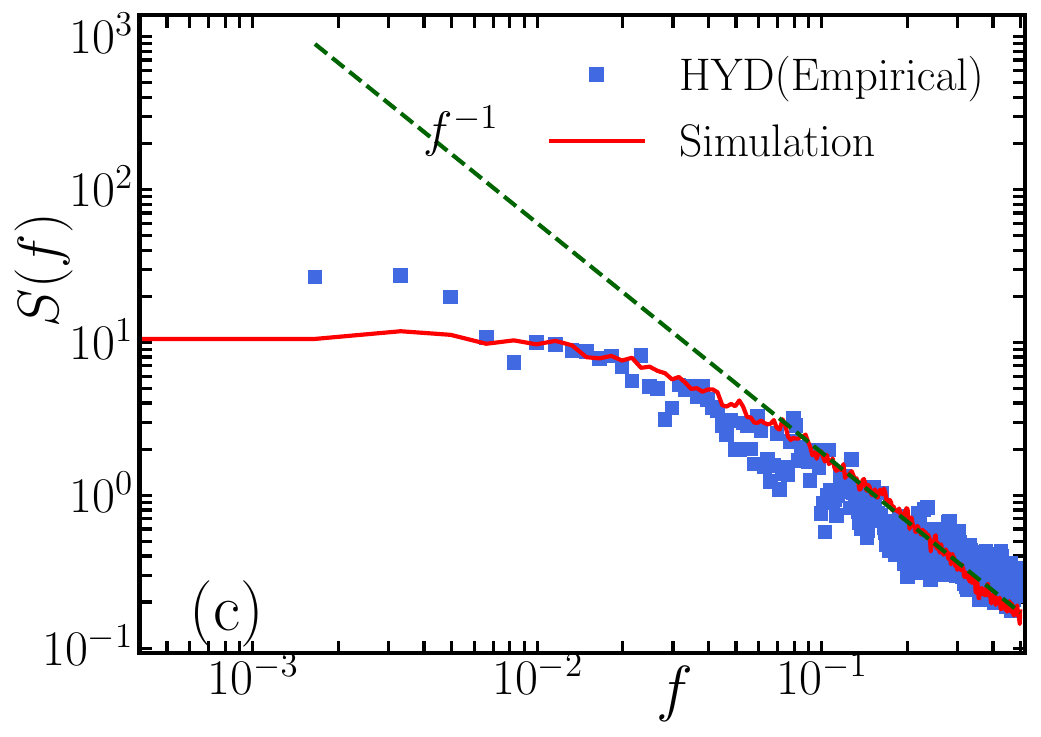}
    \caption{Plot of power spectral density $S(f)$ vs $f$ for (a) BLR, (b) KOL, and (c) HYD. The symbols indicate the $S(f)$ extracted from the empirical data while the solid lines indicate the data obtained from numerical simulations. The dashed line indicates the $1/f$ behaviour. The numerical simulations are done using the values of $\nu_\text{min}$ and $\nu_\text{max}$ estimated from the empirical data, which are given by $\nu_\text{min}=0.075$ and $\nu_\text{max}=2.0$ for BLR; $\nu_\text{min}=0.07$ and $\nu_\text{max}=2.0$ for KOL, and $\nu_\text{min}=0.056$ and $\nu_\text{max}=0.35$  for HYD.}
  \label{fig:psdsimulation}
\end{figure*}
%%%%%%%%%%%%%%%%%%%%%%%%%%%%%%%%%%%%%%%%%%%%%%%%%%%

Our objective is to recover the characteristic  $1/f$  decay of the power spectral density from the above equation, which can be achieved through the following approach. Let us consider that the \(\nu_i\) are distributed over a range $\nu_\text{min} \le \nu \le \nu_\text{max}$ with the density function,
\begin{equation}
g(\nu)=\frac{1}{\nu\,\ln\!\left(\nu_{\max}/\nu_{\min}\right)}, \label{eq:g_nu}
\end{equation} 
such that $\nu_1=\nu_\text{min}$ and $\nu_K=\nu_\text{max}$. Moreover, we assume that the weight $w_i$ of each mode is equal, i.e., $w_i= 1/\sqrt{K}$. Thus, in the large $K$ limit, the sum in Eq.~\eqref{eq:psd} can be converted to an integral, 
\begin{align}
S(f) &\approx 2 \sigma^2 \int_{\nu_{\min}}^{\nu_{\max}}  d\nu \, g(\nu) \frac{\nu}{\nu^2+4\pi^2 f^2} \cr 
& = \frac{2\sigma^2}{\ln\!\left(\nu_{\max}/\nu_{\min}\right)}
\int_{\nu_{\min}}^{\nu_{\max}}
\frac{d\nu}{\nu^2+4\pi^2 f^2}.
\end{align}
This integral can be performed exactly, yielding,
\begin{equation}
\begin{split}
S(f)\approx\;&
\frac{\sigma^2}{\pi f \ln\!\left(\nu_{\max}/\nu_{\min}\right)}
\left[
\tan^{-1}\!\left(\frac{\nu_{\max}}{2\pi f}\right)
-
\tan^{-1}\!\left(\frac{\nu_{\min}}{2\pi f}\right)
\right].
\end{split}
\end{equation}
Now, in the intermediate frequency regime, $\nu_{\min}\ll 2\pi f \ll \nu_{\max}$, we have, $\tan^{-1}\!\left(\frac{\nu_{\max}}{2\pi f}\right)\approx \frac{\pi}{2}$ and $\tan^{-1}\!\left(\frac{\nu_{\min}}{2\pi f}\right)\approx 0$. Consequently, in this regime, the PSD is expected to show the desired behaviour,
\begin{equation}\label{eq:26}
S(f)\approx
\frac{\sigma^2}{2\,\ln\!\left(\nu_{\max}/\nu_{\min}\right)}\,\frac{1}{f}.
\end{equation}

We can also compute the auto-correlation using the same approach. Using Eq.~\eqref{eq:g_nu} in Eq.~\eqref{eq:C_tau_th}, we can write, in the large $K$ limit,
\begin{align}
C(\tau) \approx \sigma^2 \int_{\nu_\text{min}}^{\nu_\text{max}} d \nu \, g(\nu) e^{-\nu \tau} 
\end{align}
Performing the integral, we get,
\begin{align}
C(\tau) \approx \sigma^2 [\text{Ei}(\nu_\text{max} \tau) -  \text{Ei}(\nu_\text{min} \tau) ], \label{eq:c_tau_theory}
\end{align}
where $\text{Ei(z)}$ denotes the exponential integral function [see Eq.(6.2.1) in Ref.~\cite{NIST:DLMF}]. The above form of auto-correlation suggests that there are two time-scales $\nu_\text{min}^{-1}$ and $\nu_\text{max}^{-1}$ associated with it. Its behaviour for large $\tau$ is determined by the longer time-scale, and is given by an exponential-like decay $C(\tau) \sim e^{-\nu_\text{min} \tau}/\tau$. On the other hand, $C(\tau)$ approaches a constant value for small $\tau$, which is consistent with the $1/f$ decay of the PSD.

We compare the auto-correlation obtained from the empirical data for three different representative cities with Eq.~\eqref{eq:c_tau_theory} in Fig.~\ref{fig:acfsimulation} where $\nu_\text{min}$ and $\nu_\text{max}$ are treated as city-dependent fitting parameters. The theoretical prediction matches with the data reasonably well. In all these cases, $1/\nu_\text{min}$ turns out to be $\sim 10$ days, explaining the observed typical decay scale of the auto-correlation function.

Thus, the minimal model proposed in Eq.~\eqref{eq:X_decomp} describes all  the universal features observed in the daily PM\textsubscript{2.5} data, namely, the EMG-like stationary distribution and the $1/f$ decay of the PSD as well as the qualitative behaviour of the auto-correlation. It should be emphasized that $\nu_\text{min}, \nu_\text{max}$ remain as non-universal free parameters. 

We next validate the model against the empirical observations through numerical simulations of the process ${\cal R}(t)$ defined in Eq.~\eqref{eq:X_decomp}. To simulate $G(t)$ we take $K=160$ OU processes which are discretized using Euler–Maruyama scheme with $dt=0.01$. The corresponding $\{\nu_i\}$ values are drawn from the distribution  $g(\nu)$ in Eq.~\eqref{eq:g_nu}. The choice of values of $\nu_\text{min}$ and $\nu_\text{max}$ is guided by the fit shown in Fig.~\ref{fig:acfsimulation}. The process $Z(t)$ is simulated by adding an exponentially distributed random number at discrete intervals $\Delta t=1$. We use the previously estimated values of $\mu, \lambda$ and $\sigma$ from the empirical data. 

From the numerical simulations, we first measure the stationary probability distribution of ${\cal R}$ using $1000$ independent realizations. Figure~\ref{fig:pdfsimulation} compares the resulting PDF with the universal scaling curve obtained from the data as well as the EMG function \eqref{eq:emg_f}, which shows excellent agreement. Note that, as mentioned already, the stationary distribution is independent of the values of $\nu_\text{min}$ and $\nu_\text{max}$. Next, we measure the auto-correlation function $C(\tau)$ and  compare this directly with the empirical data. To this end, we use the value of $\nu_\text{min}$ and $\nu_\text{max}$ obtained from fitting the same with the analytical prediction in Eq.~\eqref{eq:c_tau_theory}. This is illustrated in Fig.~\ref{fig:acfsimulation} for three representative cities. Finally, we also extract the PSD from the numerical simulations for these cities, which is shown in Fig.~\ref{fig:psdsimulation}. Clearly, the simulation data shows the $1/f$ decay at the tail, in good agreement with the empirical data. 
These results establish that the proposed stochastic framework quantitatively captures the principal empirical signatures of the system, thereby substantiating its validity as a minimal yet effective description of the observed dynamics.

\section{Conclusion}\label{sec8}

In this work we characterize the fluctuations of PM$_{2.5}$ using daily data for a period of six years from fifty four Indian cities. 
We find that, the probability density functions of the rescaled  data for most of the cities collapse onto a single curve, indicating  a universal statistical structure across locations.
To probe the intrinsic stochastic fluctuations, we exclusively focus on the residual component, which is extracted from the daily time-series after removing the long-term trend and seasonal components. 
In particular, we analyze the stationary distribution, temporal correlation and power-spectral density of the residual fluctuations. 

One of our main findings is that, despite substantial heterogeneity in local conditions, the residual fluctuations of PM$_{2.5}$ show certain universal statistical behaviour across most of the cities. This surprising universality appears in the static properties, namely, the stationary distribution of PM$_{2.5}$, as well as in dynamical features like auto-correlation and power-spectral density. First, we show that, the residual distributions of fifty-one cities converge onto a single scaling curve that is accurately captured by an exponentially modified Gaussian function. Moreover, we show that the auto-correlation function of PM$_{2.5}$ show similar behaviour, with a gradual decay at short time lags, followed by a more rapid decline. Correspondingly, the PSD shows a $1/f$ decay across all the cities.

To understand these universal features from a theoretical perspective, we propose a minimal stochastic model for the time-evolution of the PM\textsubscript{2.5}. The model describes the underlying dynamics through a superposition of Ornstein–Uhlenbeck processes, representing typical Gaussian fluctuations, combined with intermittent random kicks that introduce non-Gaussian variability. We show that this model accurately describes the stationary distribution with parameters estimated from the empirical data. Moreover, this model also explains the $1/f$ decay of the PSD along with the qualitative shape of the auto-correlation function.

Overall, these findings suggest that while urban air pollution dynamics are shaped by complex and location-specific factors, they are nonetheless constrained by underlying universal stochastic mechanisms. The modeling framework introduced here not only accounts for the observed scaling behavior but also provides a mechanistic perspective on fluctuation dynamics in environmental systems. Future work may extend this approach to incorporate nonstationary effects, explicit coupling to meteorological variables, or other pollutants, potentially revealing broader universality in urban environmental fluctuations.

\section{Data Availability}
All  data used in this work are publicly available~\cite{cpcb}.

\begin{acknowledgments}
KG acknowledges the support provided by the Interdisciplinary Statistical Research Unit and Sampling and Official Statistics Unit, Indian Statistical Institute, Kolkata during this project. \end{acknowledgments}

\appendix 

\section{Time Series Decomposition}\label{ap:decomp}

\begin{table*}[tbh]
\small
\centering
\resizebox{\textwidth}{!}{%
\begin{tabular}{|l|c|c|l|c|c|l|c|c|}
\hline
City & $N_{\alpha}$ &
\makecell{Population\\(in million)} &
City & $N_{\alpha}$ &
\makecell{Population\\(in million)} &
City & $N_{\alpha}$ &
\makecell{Population\\(in million)} \\
\hline\hline
Delhi (DEL) & 38 & 34.67 & Srinagar (SXR) & 1 & 1.78 & Mangaluru (MAN) & 1 & 0.78\\ \hline
Mumbai (MUM) & 18 & 22.09 & Agra (AGR) & 5 & 1.59 & Agartala (AGT) & 2 & 0.67\\ \hline
Kolkata (KOL) & 7 & 15.85 & Jabalpur (JBL) & 4 & 1.58 & Rajahmundry (RAJ) & 1 & 0.59\\ \hline
Bengaluru (BLR) & 7 & 14.40 & Asansol (ASN) & 1 & 1.57 & Ajmer (AJM) & 1 & 0.54 \\ \hline
Chennai (CHN) & 8 & 12.34 & Nashik (NAS) & 1 & 1.49 & Bhagalpur (BGP) & 2 & 0.53\\ \hline
Hyderabad (HYD) & 14 & 11.34 & Bhilai (BHI) & 2 & 1.32 & Shillong (SHL) & 1 & 0.52  \\ \hline
Ahmedabad (AHM) & 9 & 9.06 & Chandigarh (CHD) & 3 & 1.27 & Siliguri (SIL) & 1 & 0.51\\ \hline
Surat (SUR) & 1 & 8.58 & Guwahati (GWH) & 4 & 1.22  & Vellore (VEL) & 1 & 0.49\\ \hline
Jaipur (JAI) & 6 & 4.41 & Amritsar (AMR) & 1 & 1.13 & Tirupati (TIR) & 1 & 0.37\\ \hline
Lucknow (LKO) & 7 & 4.13 & Prayagraj (PRA) & 3 & 1.11 & Bilaspur (BIL) & 1 & 0.33\\ \hline
Indore (IND) & 1& 3.48 & Gwalior (GWA) & 1 & 1.10 & Bathinda (BHA) & 1 & 0.32\\ \hline
Kanpur (KNP) & 4 & 3.35 & Visakhapatnam (VTZ) & 1 & 1.06 & Thrissur (THR) & 1 & 0.32\\ \hline
Coimbatore (COI) & 1 & 3.16 & Dehradun (DHR) & 1 & 1.04 & Alwar (ALW) & 1 & 0.32\\ \hline
Thiruvananthapuram (TVM) & 2 & 3.07 & Jodhpur (JOD) & 1 & 1.03 & Aizawl (AIZ) & 1 & 0.29\\ \hline
Patna (PAT) & 6 & 2.69 & Kurukshetra (KUR) & 1 & 0.97 & Ratlam (RAT) & 1 & 0.27\\ \hline
Bhopal (BHP) & 3 & 2.69 & Solapur (SOL) & 1 & 0.95 & Kohima (KOH) & 1 & 0.10\\ \hline
Nagpur (NAG) & 1 & 2.41 & Puducherry (PDY) & 1 & 0.94 & Talcher (TAL) & 1 & 0.04\\ \hline
Ludhiana (LUD) & 1 & 2.03 & Salem (SAL) & 1 & 0.83 & Naharlagun (NAH) & 1 & 0.03\\ \hline
\end{tabular}
}
\caption{List of cities along with  their respective populations and corresponding number of monitoring stations $N_{\alpha}$.}
\label{city_list}
\end{table*}

\begin{table*}[tbh]
\small
\centering
% \resizebox{\textwidth}{!}{%
\begin{tabular}{|l |l |l |l |l |l |l|l|l|l|l|l|l|l|l|l|l|l|}\hline 
City& $\langle\rho\rangle$ & $s$ & City& $\langle\rho\rangle$ & $s$ & City& $\langle\rho\rangle$ & $s$ & City& $\langle\rho\rangle$ & $s$ & City& $\langle\rho\rangle$ & $s$ & City& $\langle\rho\rangle$ & $s$ \\\hline
DEL & 102.60 & 75.42 & LKO & 70.36 & 52.21 & SXR & 27.44 & 17.04 & PRA & 40.02 & 27.71 & MAN & 29.84 & 17.77 & BIL & 27.45 & 12.71\\\hline
MUM & 44.82 & 36.96 & IND & 45.37 & 25.41 & AGR & 45.25 & 39.81 & GWA & 61.06 & 48.01 & AGT & 60.58 & 36.09 & BHA & 38.28 & 27.49\\\hline
KOL & 51.30 & 41.99 & KNP & 67.62 & 49.20 & JBL & 45.45 & 30.35 & VTZ & 45.63 & 25.80 & RAJ & 33.78 & 26.11 & THR & 28.92 & 12.56\\\hline
BLR & 32.37 & 17.90 & COI & 32.16 & 12.54 & ASN & 62.69 & 34.71 & DHR & 41.05 & 24.32 & AJM & 45.16 & 17.69 & ALW & 40.78 & 15.63 \\\hline
CHN & 30.49 & 19.40 & TVM & 25.12 & 16.19 & NAS & 38.66 & 21.85 & JOD & 70.67 & 31.56 & BGP & 81.74 & 58.72 & AIZ & 13.23 & 13.87\\\hline
HYD & 37.70 & 20.32 & PAT & 86.75 & 63.25 & BHI & 25.78 & 19.26 & KUR & 64.81 & 38.56 & SHL & 18.65 & 18.29 & RAT & 45.39 & 22.64\\\hline
AHM & 49.61 & 24.38 & BHP & 46.06 & 29.61 & CHD & 58.92 & 39.69 & SOL & 37.92 & 18.74 & SIL & 47.88 & 32.93 & KOH & 31.57 & 19.88 \\\hline
SUR & 49.34 & 39.39 & NAG & 44.68 & 27.09 & GWH & 59.96 & 42.30 & PDY & 22.74 & 14.26 & VEL & 23.65 & 20.48 & TAL & 53.84 & 36.41\\\hline
JAI & 53.20 & 29.54 & LUD & 51.68 & 27.64 & AMR & 52.21 & 25.50 & SAL & 22.49 & 19.64 & TIR & 43.13 & 29.71 & NAH & 25.73 & 13.82\\\hline
\end{tabular}
\caption{Mean and standard deviation %and Entropy 
of the PM\textsubscript{2.5} concentration data.}
\label{tab:mean_sd_entropy_all}
\end{table*}

In this Appendix we provide details about the decomposition of the PM\textsubscript{2.5} time-series into the three components, as discussed in Sec~\ref{sec:uni_residual}. The observed time-series of PM\textsubscript{2.5} is represented as 
\begin{equation}
\rho_t = T_t \times S_t \times R_t,
\end{equation}
where $T_t$ denotes the long-term trend, $S_t$ the seasonal component, and $R_t$ the residual or irregular fluctuations. The decomposition proceeds by first estimating the trend component using a centered $k$-period moving average, which smooths out short-term and seasonal variations while preserving only slow, long-term changes; the trend is given by 
\begin{equation}
\widehat{T_t} = \frac{1}{k}\sum_{i=-\frac{k-1}{2}}^{\frac{k-1}{2}} \rho_{t+i},
\end{equation}
with $k=365$ since there is a presence of clear annual repetitive pattern as observed in Fig.~\ref{fig:ts}. The detrended series is then obtained by dividing the original series by the estimated trend, $D_t = \frac{\rho_t}{\widehat{T_t}} = S_t \times R_t$, which removes long-term variations while retaining seasonal and residual effects. To isolate the seasonal component, the standard ratio-to-moving-average method is applied by associating each observation with a seasonal index $d(t)$ corresponding to the day of the year, $d \in \{1,2,\ldots,365\}$, and averaging all detrended values occurring on the same calendar day across different years, $\tilde{S}_d = \frac{1}{N_d}\sum_{t:d(t)=d} D_t$, where $N_d$ is the number of available observations for that day. These preliminary seasonal factors are then normalised so that their mean over one full seasonal cycle equals unity, 
\begin{equation}
\widehat{S_t} = \frac{\tilde{S}_d}{ \frac{1}{365}\sum_{j=1}^{365} \tilde{S}_j },
\end{equation}
ensuring consistency with the multiplicative model. Finally, the residual component is extracted by removing both the trend and seasonal components from the original series, 
\begin{equation}
R_t = \frac{\rho_t}{\widehat{T_t} \times \widehat{S_t}},
\end{equation}
yielding a residual time series that represents short-term stochastic variability without any long-term trends or systematic seasonal patterns and forms the basis for subsequent statistical analysis.

\section{Exponentially Modified Gaussian Distribution}
\label{app:EMG}

An exponentially modified Gaussian distribution refers to the distribution of sum of independent Gaussian and exponentially distributed random numbers. The PDF of the EMG is quoted in Eq.~\eqref{eq:EMG} in the main text where $\mu$, $\sigma$, and $\lambda$ denote the location, scale, and rate parameters, respectively. The complementary error function is defined as
\begin{equation}
\operatorname{erfc}(u)
=
\frac{2}{\sqrt{\pi}}
\int_{u}^{\infty} e^{-t^{2}}\, dt.
\end{equation}

The mean and variance of the EMG distribution are given by
\begin{equation}
\langle \tilde{R}_t \rangle = \mu + \frac{1}{\lambda}, \quad
\langle \tilde{R}_t^2 \rangle - \langle \tilde{R}_t \rangle^2 = \sigma^2 + \frac{1}{\lambda^2}.
\end{equation}
For the empirical data considered here, the residuals are standardized such that
\begin{equation}
\langle \tilde{R}_t \rangle = 0, \qquad
\langle \tilde{R}_t^2 \rangle = 1.
\end{equation}
These constraints imply
\begin{equation}
\lambda = -\frac{1}{\mu}, \qquad \sigma^2 = 1 - \mu^2,
\end{equation}
reducing the EMG distribution to a one-parameter family given by Eq.~\eqref{eq:emg_f} in the main text, with the admissible range $-1<\mu<0$.
Finally, we note that the distribution is positively skewed with the skewness
\begin{equation}
\langle \tilde{R}_t^3 \rangle = \frac{2}{\sigma^{3}\lambda^{3}}
\left(1+\frac{1}{\sigma^{2}\lambda^{2}}\right)^{-3/2} = -2\mu^3.
\end{equation}
This provides a way to estimate the parameter $\mu$ from the skewness of the empirical data, as quoted in Eq.~\eqref{eq:mu} in the main text.  Subsequently, $\lambda$ and $\sigma$ are uniquely determined, enabling full reconstruction of the EMG distribution for comparison with empirical and simulated distributions.

\bibliography{ref}

@article{dubrulle2022multi,
title={Multi-{F}ractality, {U}niversality and {S}ingularity in {T}urbulence},
author={Dubrulle, B{\'e}rengere},
journal={Fractal and Fractional},
volume={6},
number={10},
pages={613},
year={2022},
publisher={MDPI},
doi= {10.3390/fractalfract6100613}
}

@article{donzis2020universality,
title={Universality and scaling in homogeneous compressible turbulence},
author={Donzis, Diego A and John, John Panickacheril},
journal={Physical Review Fluids},
volume={5},
number={8},
pages={084609},
year={2020},
publisher={APS},
doi={10.1103/PhysRevFluids.5.084609}
}

@article{stanley2002scale,
title     = {Scale invariance and universality in economic phenomena},
author    = {Stanley, H. Eugene and Amaral, Lu{\'i}s A. Nunes and Gopikrishnan, Parameswaran and Plerou, Vasiliki and Salinger, Michael A.},
journal   = {Journal of Physics: Condensed Matter},
volume    = {14},
number    = {9},
pages     = {2121--2151},
year      = {2002},
publisher = {IOP Publishing},
doi       = {10.1088/0953-8984/14/9/301}
}

@article{sarkar2018scaling,
title={The scaling of income distribution in {A}ustralia: Possible relationships between urban allometry, city size, and economic inequality},
author={Sarkar, Somwrita and Phibbs, Peter and Simpson, Roderick and Wasnik, Sachin},
journal={Environment and Planning B: Urban Analytics and City Science},
volume={45},
number={4},
pages={603--622},
year={2018},
publisher={SAGE Publications Sage UK: London, England},
doi       = {10.1177/0265813516676488}
}

@article{macedo2017universality,
title     = {Universality classes of fluctuation dynamics in hierarchical complex systems},
author    = {Mac{\^e}do, A. M. S. and Gonz{\'a}lez, Iv{\'a}n R. Roa and Salazar, D. S. P. and Vasconcelos, G. L.},
journal   = {Physical Review E},
volume    = {95},
number    = {3},
pages     = {032315},
year      = {2017},
publisher = {American Physical Society},
doi       = {10.1103/physreve.95.032315}
}

@article{bettencourt2020demography,
title={Demography and the emergence of universal patterns in urban systems},
author={Bettencourt, Lu{\'\i}s MA and Z{\"u}nd, Daniel},
journal={Nature Communications},
volume={11},
number={1},
pages={4584},
year={2020},
publisher={Nature Publishing Group UK London},
doi={10.1038/s41467-020-18205-1}
}

@article{Ghosh_2019,
title = {Universal {C}ity-size distributions through rank ordering},
author = {Abhik Ghosh and Banasri Basu},
journal = {Physica A: Statistical Mechanics and its Applications},
volume = {528},
pages = {121094},
year = {2019},
issn = {0378-4371},
doi = {https://doi.org/10.1016/j.physa.2019.121094},
url ={https://www.sciencedirect.com/science/article/pii/S0378437119306703}
}

@article{youn2016scaling,
title={Scaling and universality in urban economic diversification},
author={Youn, Hyejin and Bettencourt, Lu{\'\i}s MA and Lobo, Jos{\'e} and Strumsky, Deborah and Samaniego, Horacio and West, Geoffrey B},
journal={Journal of The Royal Society Interface},
volume={13},
number={114},
pages={20150937},
year={2016},
publisher={The Royal Society},
doi={10.1098/rsif.2015.0937}
}

@article{LaPorta2024,
author    = {Caterina A. M. La Porta and Stefano Zapperi},
title     = {{U}rban {S}caling {F}unctions: {E}mission, {P}ollution and {H}ealth},
journal   = {Journal of Urban Health},
year      = {2024},
volume    = {101},
pages     = {752--763},
doi       = {10.1007/s11524-024-00888-2}
}

@article{Dodds1999,
author    = {Peter Sheridan Dodds and Daniel H. Rothman},
title     = {Unified view of scaling laws for river networks},
journal   = {Physical Review E},
year      = {1999},
volume    = {59},
pages     = {4865},
doi       = {10.1103/PhysRevE.59.4865}
}

@article{smyth2019self,
title={Self-organized criticality in geophysical turbulence},
author={Smyth, William D and Nash, JD and Moum, JN},
journal={Scientific reports},
volume={9},
number={1},
pages={3747},
year={2019},
publisher={Nature Publishing Group UK London},
doi={10.1038/s41598-019-39869-w}
}

@article{Entropy2021,
title={Stochastic and {S}elf-{O}rganisation {P}atterns in a 17-year {PM}\textsubscript{10} {T}ime {S}eries in {A}thens, {G}reece},
author={Nikolopoulos, Dimitrios and Alam, Aftab and Petraki, Ermioni and Papoutsidakis, Michail and Yannakopoulos, Panayiotis and Moustris, Konstantinos P},
journal={Entropy},
volume={23},
number={3},
pages={307},
year={2021},
publisher={MDPI},
doi       = {10.3390/e23030307}
}

@article{Abbasi2024,
author    = {M.T. Abbasi and A.A. Alesheikh and A. Jafari and others},
title     = {Spatial and temporal patterns of urban air pollution in {T}ehran with a focus on {PM}\textsubscript{2.5} and associated pollutants},
journal   = {Scientific Reports},
year      = {2024},
volume    = {14},
pages     = {25150},
doi       = {10.1038/s41598-024-75678-6}
}

@article{DavidSen2007,
title={{S}caling and {U}niversality in {R}ock {F}racture},
author={Davidsen, J{\"o}rn and Stanchits, Sergei and Dresen, Georg},
journal={Physical Review Letters},
volume={98},
number={12},
pages={125502},
year={2007},
publisher={APS},
doi       = {10.1103/physrevlett.98.125502}
}

@article{Bran_2024,
title = {Understanding the seasonal dynamics of surface {PM}\textsubscript{2.5} mass distribution and source contributions over {T}hailand},
journal = {Atmospheric Environment},
volume = {331},
pages = {120613},
year = {2024},
issn = {1352-2310},
doi = {10.1016/j.atmosenv.2024.120613},
url = {https://www.sciencedirect.com/science/article/pii/S1352231024002887},
author = {Sherin {Hassan Bran} and Ronald Macatangay and Chakrit Chotamonsak and Somporn Chantara and Vanisa Surapipith}
}

@article{Mishra_2021,
title = {An application of probability density function for the analysis of {PM}2.5 concentration during the {COVID}-19 lockdown period},
journal = {Science of The Total Environment},
volume = {782},
pages = {146681},
year = {2021},
issn = {0048-9697},
doi = {https://doi.org/10.1016/j.scitotenv.2021.146681},
url = {https://www.sciencedirect.com/science/article/pii/S0048969721017496},
author = {Gaurav Mishra and Kunal Ghosh and Anubhav Kumar Dwivedi and Manish Kumar and Sidyant Kumar and Sudheer Chintalapati and S.N. Tripathi}
}

@article{Beck_2020,
title = {Superstatistical approach to air pollution statistics},
author = {Williams, Griffin and Sch\"afer, Benjamin and Beck, Christian},
journal = {Phys. Rev. Res.},
volume = {2},
issue = {1},
pages = {013019},
numpages = {9},
year = {2020},
month = {Jan},
publisher = {American Physical Society},
doi = {10.1103/PhysRevResearch.2.013019}
}

@article{ShiLiuHuang2015,
author  = {Shi, Kai and Liu, Chunqiong and Huang, Yi},
title   = {{M}ultifractal {P}rocesses and {S}elf-{O}rganized {C}riticality of {PM}2.5 during a {T}ypical {H}aze {P}eriod in {C}hengdu, {C}hina},
journal = {Aerosol and Air Quality Research},
volume  = {15},
number  = {3},
pages   = {926--934},
year    = {2015},
doi     = {10.4209/aaqr.2014.05.0091}
}

@book{gibbs2011advanced,
title={Advanced {K}alman filtering, least-squares and modeling: a practical handbook},
author={Gibbs, Bruce P},
year={2011},
publisher={John Wiley \& Sons}
}

@ARTICLE{akf,
author={Kruse, Theresa and Griebel, Thomas and Graichen, Knut},
journal={IEEE Access}, 
title={Adaptive {K}alman {F}iltering: {M}easurement and {P}rocess {N}oise {C}ovariance {E}stimation {U}sing {K}alman {S}moothing}, 
year={2025},
volume={13},
number={},
pages={11863-11875},
doi={10.1109/ACCESS.2025.3528348}
}

@article{KL,
author = {S. Kullback and R. A. Leibler},
title = {{On {I}nformation and {S}ufficiency}},
volume = {22},
journal = {The Annals of Mathematical Statistics},
number = {1},
publisher = {Institute of Mathematical Statistics},
pages = {79 -- 86},
year = {1951},
doi = {10.1214/aoms/1177729694}
}

@ARTICLE{js,
author={Lin, J.},
journal={IEEE Transactions on Information Theory}, 
title={Divergence measures based on the {S}hannon entropy}, 
year={1991},
volume={37},
number={1},
pages={145-151},
doi={10.1109/18.61115}
}

@article{multiplyts,
title={Spatio-{T}emporal {PM}2.5 {F}orecasting {U}sing {M}achine {L}earning and {L}ow-{C}ost {S}ensors: {A}n {U}rban {P}erspective},
author={Zareba, Mateusz and Cogiel, Szymon and Danek, Tomasz},
journal={Engineering Proceedings},
volume={101},
number={1},
pages={6},
year={2025},
publisher={MDPI},
doi= {10.3390/engproc2025101006}
}

@book{world2021global,
  author    = {{World Health Organization}},
  title     = {WHO Global Air Quality Guidelines: Particulate Matter ({PM}\textsubscript{2.5} and {PM}\textsubscript{10}), Ozone, Nitrogen Dioxide, Sulfur Dioxide and Carbon Monoxide},
  year      = {2021},
  publisher = {World Health Organization},
  address   = {Geneva}
}

@article{Pope01062006,
author = {C. Arden Pope III and Douglas W. Dockery},
title = {{H}ealth {E}ffects of {F}ine {P}articulate {A}ir {P}ollution: {L}ines that {C}onnect},
journal = {Journal of the Air \& Waste Management Association},
volume = {56},
number = {6},
pages = {709--742},
year = {2006},
publisher = {Taylor \& Francis},
doi = {10.1080/10473289.2006.10464485}
}

@article{burnett2018global,
title={Global estimates of mortality associated with long-term exposure to outdoor fine particulate matter},
author={Burnett, Richard and Chen, Hong and Szyszkowicz, Mieczys{\l}aw and Fann, Neal and Hubbell, Bryan and Pope III, C Arden and Apte, Joshua S and Brauer, Michael and Cohen, Aaron and Weichenthal, Scott and others},
journal={Proceedings of the National Academy of Sciences},
volume={115},
number={38},
pages={9592--9597},
year={2018},
publisher={National Academy of Sciences},
doi={10.1073/pnas.1803222115}
}

@book{seinfeld_2016,
  title={Atmospheric Chemistry and Physics: From air pollution to climate change},
  author={Seinfeld, John H and Pandis, Spyros N},
  year={2016},
  publisher={John Wiley \& Sons}
}

@article{Kelp_2022,
author = {Kelp, Makoto M and Lin, Samuel and Kutz, J Nathan and Mickley, Loretta J},
title = {A new approach for determining optimal placement of {PM}\textsubscript{2.5} air quality sensors: Case study for the contiguous {U}nited {S}tates},
year = {2022},
publisher = {IOP Publishing},
volume = {17},
number = {3},
pages = {034034},
journal = {Environmental Research Letters},
doi = {10.1088/1748-9326/ac548f}
}

@article{He_2022,
    author = {He, H and Schafer, B and Beck, C},
    title = {Spatial heterogeneity of air pollution statistics in {E}urope},
    journal = {Scientific Reports},
    volume = {12},
    number = {1},
    pages = {12215},
    year = {2022},
    doi = {doi: 10.1038/s41598-022-16109-2}
}

@misc{cpcb,
  author       = {{Central Pollution Control Board (CPCB)}},
  title        = {National Air Monitoring Programme (NAMP) Data},
  year         = {2025},
  url          = {https://cpcb.nic.in/namp-data/},
  note         = {Accessed: 2025-07-10}
}

@article{Eliazar_2010,
  title = {Universal generation of $1/f$ noises},
  author = {Eliazar, Iddo and Klafter, Joseph},
  journal = {Phys. Rev. E},
  volume = {82},
  issue = {2},
  pages = {021109},
  numpages = {8},
  year = {2010},
  month = {Aug},
  publisher = {American Physical Society},
  doi = {10.1103/PhysRevE.82.021109},
  url = {https://link.aps.org/doi/10.1103/PhysRevE.82.021109}
}

@article{Yamamoto_2014,
  title = {Stochastic model of {Z}ipf's law and the universality of the power-law exponent},
  author = {Yamamoto, Ken},
  journal = {Phys. Rev. E},
  volume = {89},
  issue = {4},
  pages = {042115},
  numpages = {4},
  year = {2014},
  month = {Apr},
  publisher = {American Physical Society},
  doi = {10.1103/PhysRevE.89.042115},
  url = {https://link.aps.org/doi/10.1103/PhysRevE.89.042115}
}

@article{iyerbiswas_2014,
  title = {{U}niversality in {S}tochastic {E}xponential {G}rowth},
  author = {Iyer-Biswas, Srividya and Crooks, Gavin E. and Scherer, Norbert F. and Dinner, Aaron R.},
  journal = {Phys. Rev. Lett.},
  volume = {113},
  issue = {2},
  pages = {028101},
  numpages = {5},
  year = {2014},
  month = {Jul},
  publisher = {American Physical Society},
  doi = {10.1103/PhysRevLett.113.028101},
  url = {https://link.aps.org/doi/10.1103/PhysRevLett.113.028101}
}

@article{Lowen_1993,
  title = {Fractal renewal processes generate 1/f noise},
  author = {Lowen, S. B. and Teich, M. C.},
  journal = {Phys. Rev. E},
  volume = {47},
  issue = {2},
  pages = {992--1001},
  numpages = {0},
  year = {1993},
  month = {Feb},
  publisher = {American Physical Society},
  doi = {10.1103/PhysRevE.47.992},
  url = {https://link.aps.org/doi/10.1103/PhysRevE.47.992}
}

@article{ali2022comparison,
  title={A comparison of different parameter estimation methods for exponentially modified {G}aussian distribution},
  author={Ali, Sajid and Ara, Jehan and Shah, Ismail},
  journal={Afrika Matematika},
  volume={33},
  number={2},
  pages={58},
  year={2022},
  publisher={Springer},
  doi={10.1007/s13370-022-00995-w}
}

@article{Zhen_2014,
title = {Exponentially modified {G}aussian relevance to the distributions of translocation events in nanopore-based single molecule detection},
journal = {Chinese Chemical Letters},
volume = {25},
number = {7},
pages = {1029-1032},
year = {2014},
issn = {1001-8417},
doi = {https://doi.org/10.1016/j.cclet.2014.05.009},
url = {https://www.sciencedirect.com/science/article/pii/S1001841714002186},
author = {Zhen Gu and Yi-Lun Ying and Bing-Yong Yan and Hui-Feng Wang and Pin-Gang He and Yi-Tao Long},
}

@article{Golubev_2010,
title = {Exponentially modified {G}aussian ({EMG}) relevance to distributions related to cell proliferation and differentiation},
journal = {Journal of Theoretical Biology},
volume = {262},
number = {2},
pages = {257-266},
year = {2010},
issn = {0022-5193},
doi = {https://doi.org/10.1016/j.jtbi.2009.10.005},
url = {https://www.sciencedirect.com/science/article/pii/S0022519309004809},
author = {A. Golubev}}

@article{rmse1,
  title={A comparative study of imputation techniques for missing values in healthcare diagnostic datasets},
  author={Joel, Luke Oluwaseye and Doorsamy, Wesley and Paul, Babu Sena},
  journal={International Journal of Data Science and Analytics},
  pages={1},
  year={2025},
  publisher={Springer},
  doi={10.1007/s41060-025-00825-9}
}

@incollection{nielsen2016hierarchical,
  title={Hierarchical clustering},
  author={F. Nielsen},
  booktitle={Introduction to HPC with MPI for Data Science},
  pages={195--211},
  year={2016},
  publisher={Springer}
}

@article{hierarchical_clustering1,
author={L. Zhang and G. Yang},
title = {Cluster analysis of {PM}\textsubscript{2.5} pollution in {C}hina using the frequent itemset clustering approach},
journal = {Environmental Research},
volume = {204},
pages = {112009},
year = {2022},
issn = {0013-9351},
url = {https://doi.org/10.1016/j.envres.2021.112009}
}

@misc{airnow_aqi_basics,
  author       = {{AirNow}},
  title        = {AQI {B}asics},
  url          = {https://www.airnow.gov/aqi/aqi-basics/},
  note         = {Accessed: 2026-03-24}
}

@article{KG_2026,
    author = {Ghosh, Koyena and Banerjee, Suchismita and Basu, Urna and Basu, Banasri},
    title = {Classifying urban regions by aggregated pollutant–weather correlation strength: a spatiotemporal study},
    volume = {40},
    journal = {Stochastic Environmental Research and Risk Assessment},
    number = {76},
    issue = {4},
    pages = {1436-3259},
    year = {2026},
    doi = {10.1007/s00477-026-03193-3},
    url = {https://doi.org/10.1007/s00477-026-03193-3}
}

@article{WK_theorem1,
  author={N. Wiener},
  title={{Generalized harmonic analysis}},
  volume={55},
  journal={Acta Mathematica},
  number={1},
  publisher={Springer},
  pages={117--258},
  year={1930},
  doi = {10.1007/BF02546511}
}

@article{WK_theorem2,
  author={A. Khinchin},
  title={{Korrelationstheorie der station{\"a}ren stochastischen Prozesse}},
  volume={109},
  journal={Mathematische Annalen},
  number={1},
  publisher={Springer},
  pages={604--615},
  year={1934},
  doi = {https://doi.org/10.1007/BF01449156}
}

@article{welch,
  author={P. D. Welch},
  title={{The use of fast Fourier transform for the estimation of power spectra: A method based on time averaging over short, modified periodograms}},
  volume={15},
  journal={IEEE Transactions on Audio and Electroacoustics},
  number={2},
  pages={70--73},
  year={1967},
  doi = {10.1109/TAU.1967.1161901}
}

@article{Keshner_1982,
  title={1/f noise},
  author={Marvin S. Keshner},
  journal={Proceedings of the IEEE},
  year={1982},
  volume={70},
  pages={212-218},
  doi = {10.1109/PROC.1982.12282},
  url={https://api.semanticscholar.org/CorpusID:921772}
}

@article{Weissman_1988,
  title={1/f noise and other slow, nonexponential kinetics in condensed matter.},
  author={M. B. Weissman},
  journal={Reviews of Modern Physics},
  year={1988},
  volume={60},
  pages={537-571},
  doi = {10.1103/REVMODPHYS.60.537},
  url={https://api.semanticscholar.org/CorpusID:120723502}
}

@misc{NIST:DLMF,
         key = "{\relax DLMF}",
       title = "{\it NIST Digital Library of Mathematical Functions}",
howpublished = "\url{https://dlmf.nist.gov/}, Release 1.2.6 of 2026-03-15",
         url = "https://dlmf.nist.gov/",
        note = "F.~W.~J. Olver, A.~B. {Olde Daalhuis}, D.~W. Lozier, B.~I. Schneider,
                R.~F. Boisvert, C.~W. Clark, B.~R. Miller, B.~V. Saunders,
                H.~S. Cohl, and M.~A. McClain, eds."}
\end{document}